\documentclass{elsart}
\journal{New Astronomy}
\usepackage{natbib}

\usepackage{graphicx}
\usepackage{amssymb}

\def\astrobj#1{#1}
\def\url#1{{\ttfamily\def\/{/\discretionary{}{}{}}#1}}
\def\bibcode#1{}%invisible in printed text

\begin{document}

\begin{frontmatter}
\title{Scalar potential model of redshift and discrete redshift }
\author{John C. Hodge}%\thanksref{a}}
\address{ Blue Ridge Community College, 100 College Dr., Flat Rock, NC, 28731-1690}
%\thanks[a]{Visiting from XZD Corp., 3 Fairway St., Brevard, NC, 28712, email: %scjh@citcom.net }
% use the thanksref command within \title, \author or \address for footnotes:
% \title{\thanksref{label1}}
% \thanks[label1]{}
% \author{\thanksref{label2}}
% \thanks[label2]{}
% \address{\thanksref{label3}}
% \thanks[label3]{}
% including your email address
% \address{\thanksref{email}}
\thanks[email]{E-mail: jch9496@blueridge.edu }

\begin{abstract}
On the galactic scale the universe is inhomogeneous and redshift $z$ is occasionally less than zero.  A scalar potential model (SPM) that links the galaxy scale $z$ to the cosmological scale $z$ of the Hubble Law is postulated.  Several differences among galaxy types suggest that spiral galaxies are Sources and that early type, lenticular, and irregular galaxies are Sinks of a scalar potential field.  The morphology-radius and the intragalactic medium cluster observations support the movement of matter from Source galaxies to Sink galaxies.  A cell structure of galaxy groups and clusters is proposed to resolve a paradox concerning the scalar potential like the Olber's paradox concerning light.  For the sample galaxies, the ratio of the luminosity of Source galaxies to the luminosity of Sink galaxies approaches $2.7 \pm 0.1$.  An equation is derived from sample data, which is anisotropic and inhomogeneous, relating $z$ of and the distance $D$ to galaxies.  The calculated $z$ has a correlation coefficient of 0.88 with the measured $z$ for a sample of 32 spiral galaxies with $D$ calculated using Cepheid variable stars.  The equation is consistent with $z<0$ observations of close galaxies.  At low cosmological distances, the equation reduces to $z \approx \exp(KD) \, -1 \approx KD$, where $K$ is a constant, positive value.  The equation predicts $z$ from galaxies over 18 Gpc distant approaches a constant value on the order of 500.  The SPM of $z$ provides a physical basis for the $z$ of particle photons.  Further, the SPM qualitatively suggests the discrete variations in $z$, which was reported by W. G. Tifft, 1997, Astrophy. J. 485, 465 and confirmed by others, are consistent with the SPM.
\end{abstract}

\end{frontmatter}

% main text
\section{INTRODUCTION}

The evidence suggest the problem of a single model explaining both galactic scale and cosmological scale observations is fundamental~\citep{sell}.  Among a variety of models that have been suggested to link cosmological scale and galactic scale observations are models using a scalar field.  A scalar field has been linked to dark matter \citep{fay,piro}, a cosmological model \citep{agui}, the rotational curves of spiral galaxies \citep{mbel}, and axisymmetric galaxies \citep{rodr}.

The great majority of elliptical galaxies are observed to be much poorer in cool gas and hydrogen than spiral galaxies of comparable luminosity \citep[pages 527-8]{binn}.  The bulk of the interstellar matter (ISM) in spiral galaxies is H{\scriptsize{I}} and hydrogen.  In elliptical galaxies, the bulk of the ISM consists of hot plasma distributed approximately spherically rather than in a thin disk \citep[pages 525-6]{binn}.  A characteristic of elliptical galaxies not found in spiral galaxies is that the X-ray surface brightness is nearly proportional to the optical surface brightness \citep[pages 526]{binn}.  The study of dust lanes suggests that gas and dust are falling into elliptical and lenticular galaxies \citep[pages 513-6]{binn} and are formed internally in spiral galaxies \citep[pages 528-9]{binn}.  Some evidence has been presented that suggests irregular galaxies will settle down to being a normal elliptical galaxy \citep[page 243]{binn}.  In low surface brightness (LSB) spiral galaxies, the outer rotation curve (RC) generally rises \citep[and references therein]{debl}.  In contrast, ``ordinary'' elliptical galaxies, with luminosities close to the characteristic $L^*$ (=2.2 $\times 10^{10} \, L_{B,\odot}$ in B band solar units for a Hubble constant $H_\mathrm{o} = 70 $ km~s$^{-1}$~Mpc$^{-1}$) show a nearly Keplerian decline with radius outside $2R_\mathrm{eff}$, where $R_\mathrm{eff}$ is the galaxy's ``effective radius'' enclosing half its projected light \citep{roma}.

\citet{batt} and \citet{sofu} provides an overview of the current state of knowledge of RCs of spiral galaxies.  The RCs of spiral galaxies have a high rotation velocity of over 1000 km~s$^{-1}$ near the nucleus \citep{ghez,taka}.  \citet{ghez} and \citet{fer5} have observed Keplerian motion to within one part in 100 in elliptical orbits of stars that are from a few 100 pc to a few 1000 pc from the center of spiral galaxies.  The Keplerian characteristic decline in rotation velocity is sometimes seen in H$_\mathrm{\alpha}$ RCs.  This is followed by gradual rise to a knee or sharp change of slope at a rotation velocity of less than 300 km~s$^{-1}$.  The outer RC is beyond the knee.  Interacting galaxies often show perturbed outer RCs.  The outer part of an RC is often relatively flat with rising and falling RCs occasionally being found.  

The particles most often measured in the disk region of a galaxy are hydrogen gas by H{\scriptsize{I}} observation and stars by observing the H$_\mathrm{\alpha}$ line.  The particles being measured in the inner bulge region are stars by observation of H$_\mathrm{\alpha}$, CO, and other spectral lines.  Also, the RC differs for different particles.  For example, the H{\scriptsize{I}} and H$_\mathrm{\alpha}$ RCs for \astrobj{NGC~4321} \citep{semp} differ in the outer bulge and approach each other in the outer disk region.  

\citet{mclu} found that the mass in the central region of spiral galaxies is 0.0012 of the mass of the bulge.  \citet{fer5} reported that about 0.1\% of a spiral galaxy's luminous mass is at the center of galaxies and that the density of supermassive black holes SBH's in the universe agrees with the density inferred from observation of quasars.  \citet{merr} found similar results in their study of the relationship of the mass $M_\mathrm{\bullet}$ of the central supermassive black hole and the velocity dispersion $\sigma_\mathrm{v}$ ($M_\mathrm{\bullet} - \sigma_\mathrm{v}$ relation).  \citet{fer4} found a tight relation between rotation velocity $v_\mathrm{c}$ in the outer disk region and bulge velocity dispersion $\sigma_\mathrm{c}$ ($v_\mathrm{c} - \sigma_\mathrm{c}$) which is strongly supporting a relationship of a center force with total gravitational mass of a galaxy.  \citet{wand} showed $M_\mathrm{\bullet}$ of AGN galaxies and their bulge luminosity follow the same relationships as their ordinary (inactive) galaxies, with the exception of narrow line AGN.  \citet{grah,grah2,grah3} found correlations between $M_\mathrm{\bullet}$ and structural parameters of elliptical galaxies and bulges.  Either the dynamics of many spiral galaxies are producing the same increase of mass in the center at the same rate, a feedback controlled mechanism exists to evaporate the mass increase that changes as the rate of inflow changes as suggested by \citet{merr2}, or sufficient mass is ejected from spiral galaxies to maintain the correlations.  The RCs imply the dynamics of spiral galaxies differ so the former is unlikely.  The study of dust lanes suggests mass is ejected from spiral galaxies.

The following is taken from \citet{binn}.  Galaxies in groups and clusters (``clusters'') are much more likely to be elliptical or lenticular than in the field.  Spiral galaxies lie farther from the center of clusters than do elliptical galaxies.  The fraction $f(E)$ of galaxies that are elliptical galaxies in clusters varies from 15\% to 40\%.  Clusters with a large value of $f(E)$ tend to have a regular, symmetric appearance, often with a large cD galaxy at its center.  Clusters with a low value of $f(E)$ generally have a ratty appearance.  The fraction $f(Sp)$ of spiral galaxies in centrally-concentrated clusters increases with radius $R$ from the center of the cluster.  The observations are consistent with the model of their being no spiral galaxies in the cores of regular clusters.  The lenticular (S0) galaxies become increasingly dominant at small radii.  Nearer the core the fraction $f(S0)$ of S0 galaxies declines sharply as $f(E)$ increases sharply.  Also, the $f(E)$ increases and $f(Sp)$ decreases as the projected number $N_\mathrm{d}$ density of galaxies increases.  There appears to be a close relation between $N_\mathrm{d}$ and $R$.  The morphology of galaxies appears strongly correlated with the current surrounding density.  A galaxies radius within a cluster appears to be the primary factor that dictates its morphology except where a spiral galaxy has a nearby companion.  The nearby companion has a disproportionate probability of being elliptical.  Studies of H{\scriptsize{I}} in spiral galaxies show the average H{\scriptsize{I}} deficiency varies monotonically with $R$.  Most rich clusters of galaxies are filled with plasma with temperature $> 10^7$ K (intracluster medium).  The intracluster medium mass exceeds the total mass in the luminous parts of cluster galaxies.  The intracluster medium is composed of hydrogen-helium plasma and significant numbers of heavy-element ions, which were probably formed in galaxies.  The hot plasma may be condensing into galaxies (S0 and E) that are found at the center of clusters.  

That the redshift $z$ of emitted photons from galaxies generally increases with the extragalactic distance $D$ (Mpc) between the emitting galaxy and observer is well known ($z - D$ relationship).  Currently fashionable, cosmological models attributes $z$ to a Doppler shift of a light wave.  The assumption of homogeneity in cosmological scale volumes is compatible with either a static galaxy distribution or with a very special velocity field obeying the Hubble Law \citep[page~396]{binn}
\begin{equation}
D =  \frac{c}{H_\mathrm{o}}z
\label{eq:1},
\end{equation}
where $H_\mathrm{o}$ (km~s$^{-1}$~Mpc$^{-1}$) is the Hubble constant and $c$ (km~s$^{-1}$) is the speed of light.  The $H_\mathrm{o}$ occupies a pivotal role in current cosmologies.  The methods of calculating supernova distances, the cosmological microwave background (CMB) power spectrum, weak gravitational lensing, cluster counts, baryon oscillation, expansion of the universe, and the fundamental aspects of the Big Bang model depend on $H_\mathrm{o}$ and the Hubble law.

\begin{figure}
\begin{center}
\includegraphics[width=0.8\textwidth]{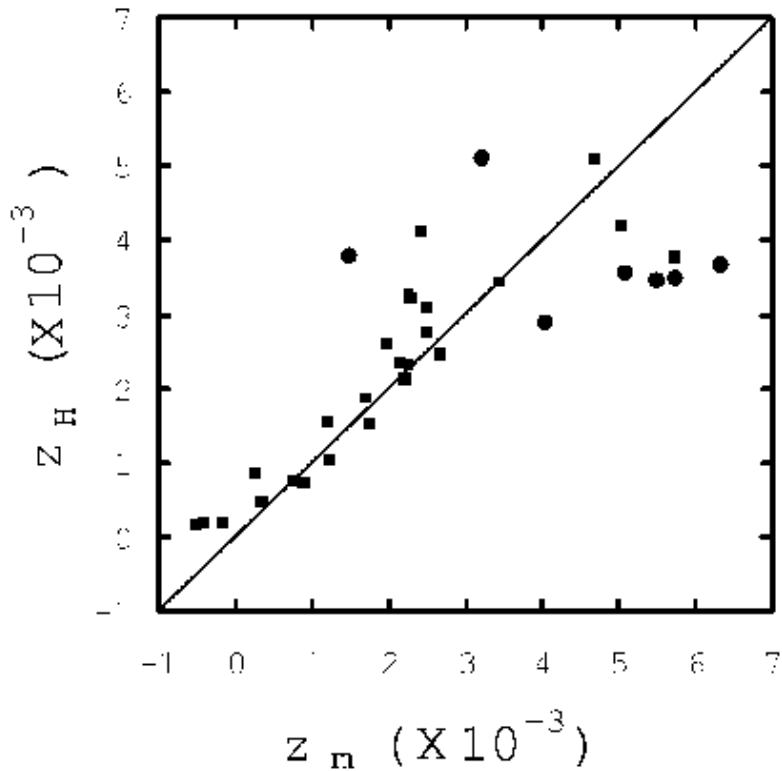}
\end{center}
\caption{Plot of the calculated redshift $z_\mathrm{H}$ using Eq.~(\ref{eq:1}) and $D$ calculated using Cepheid variable stars for 32 galaxies \citep{free,macr} versus the measured redshift $z_\mathrm{m}$.  The straight line is a plot of $z_\mathrm{H} = z_\mathrm{m}$.  The circles indicate the data points for galaxies with (l,b) = (290$^\circ \pm 20^\circ$,75$^\circ \pm 15^\circ$).}
\label{fig:1}
\end{figure}

However, the determination of $H_\mathrm{o}$ has a large uncertainty and different researchers calculate different values.  Figure~\ref{fig:1} shows the calculated redshift $z_\mathrm{H}$ using Eq.~(\ref{eq:1}), $H_\mathrm{o}=70$ km~s$^{-1}$~Mpc$^{-1}$, and the distance $D_\mathrm{a}$ calculated using Cepheid variable stars for 32 galaxies \citep{free, macr} versus the measured galactocentric redshift $z_\mathrm{m}$.  The correlation coefficient of $z_\mathrm{H}$ versus $z_\mathrm{m}$ is 0.80.

The deviation of the recession velocity of a galaxy from the straight line of the Hubble Law is ascribed to a ``peculiar velocity'' of the photon emitting galaxy relative to earth \citep[page 439]{binn}.  The deviation from the Hubble Law is the only means of determining the peculiar velocity of a galaxy.  The average peculiar velocity for all galaxies is assumed to be zero on the scale that the universe appears homogenous.  The 2dFGRS \citep{peac} suggests this scale is $z > 0.2$.

The circles in Fig.~\ref{fig:1} denote data for galaxies in the general direction of $( l, b)=(290^\circ \pm 20^\circ, 75^\circ \pm 15^\circ)$.  \citet{aaro2} found the peculiar velocity field in the local supercluster is directed toward \astrobj{NGC~4486} (\astrobj{Messier~087}) with $( l, b) \approx (284^\circ, 74^\circ)$ at a speed of 331$\pm$41 km~s$^{-1}$.  This has been called the ``Virgocentric infall''.  \astrobj{NGC~4486} is a peculiar, large, elliptical galaxy with strong X-ray emissions.  In addition, \citet{lilj} detected a quadrupolar tidal velocity field from spiral galaxy data in addition to the Virgocentric infall pointing toward (l,b) = (308$^\circ \, \pm 13^\circ $, 13$^\circ \, \pm 9^\circ $) at a speed of $\approx 200$ km~s$^{-1}$.  \astrobj{NGC~5128} (\astrobj{Centaurus A}) at (l,b) = (310$^\circ$, 19$^\circ$) and at a distance of $ 3.84 \pm 0.35$~Mpc \citep{rejk} is a galaxy with properties \citep{isra} similar to \astrobj{NGC~4486}.  \citet{lynd} found elliptical galaxies at distances in the 2000-7000 km~s$^{-1}$ range are streaming toward a ``Great Attractor'' at $( l, b) = (307^\circ \pm13^\circ, 9^\circ \pm8^\circ)$.  \astrobj{Centaurus B} at (l,b) = (310$^\circ$, 2$^\circ$) is a galaxy with properties similar to \astrobj{NGC~4486}.  In a more recent analysis, \citet{huds} suggested a bulk flow of 225 km~s$^{-1}$ toward $( l, b) \approx (300^\circ, 10^\circ)$.  However, the total mass in these directions appears to be insufficient to account for the peculiar velocity fields using Newtonian dynamics.

\citet{burb} found a $z_\mathrm{m} $ periodicity $\Delta z_\mathrm{m}  \approx 0.06$ for approximately 70 QSOs.  \citet{duar} confirmed these claims of periodicity using various statistical tests on $z_\mathrm{m}$ of 2164 QSOs.  However, another claim \citep{karl} that $\ln ( 1 + z_\mathrm{m})$ is periodic with a period of 0.206 was found to be unsupported.  \citet[and references therein]{bell4} offered further evidence of periodic $z_\mathrm{m}$ in QSOs. 

\citet{tiff2,tiff} found ``discrete velocity periods'' of $z_\mathrm{m}$ of spiral galaxies in clusters.  The ``discrete velocity periods'' of $z_\mathrm{m}$ showed an octave, or doubling, nature.  \citet[and references therein]{bell2} confirmed the existence of the ``discrete velocity periods'' of $z_\mathrm{m} $ for 83 ScI galaxies.  \citet[and references therein]{russ} presented evidence that redshifts from normal spiral galaxies have an ``intrinsic'' component that causes the appearance of peculiar velocities in excess of 5000 km~s$^{-1}$. 

In this Paper, galaxy and cluster observations are used to derive the characteristics of a scalar potential model (SPM).  The SPM suggests spiral galaxies are Sources of the scalar potential field and early type galaxies are Sinks of the scalar potential field.  The cluster observations support the movement of matter from Source galaxies to Sink galaxies.  An equation is derived that recovers Eq.~(\ref{eq:1}) for cosmological distances from data that is anisotropic and inhomogeneous.  The resulting model is used to calculate redshift of particle photons for a sample of 32 galaxies with $D_\mathrm{a}$.  The calculated redshift $z_\mathrm{c}$ has a correlation coefficient of 0.88 with $z_\mathrm{m}$.  Further, the SPM suggests the discrete variations in $z_\mathrm{m}$ \citep{bell4,bell3,bell2,tiff2,tiff} are consistent with the SPM. 

The object of this article is to derive a SPM based on galaxy and cluster observations, to apply the SPM by developing an equation for the calculation of $z$, and to test the equation using observations of galaxies with a $D_\mathrm{a}$.  Sink and Source galaxies are discussed in Section~\ref{sec:sink}.  In section~\ref{sec:model}, the SPM $z$ calculation equation is developed.  The resulting model is used to calculate $z_\mathrm{c}$ in Section~\ref{sec:results}.  Section~\ref{sec:Xfactor} discusses the factors of the $z_\mathrm{c}$ calculation.  Section~\ref{sec:tiff} discusses the discrete variations in $z_\mathrm{m}$ relative to the model of Section~\ref{sec:Xfactor}.  The discussion and conclusion is in Section~\ref{sec:disc}.  

\section{\label{sec:sink}Sink and Source galaxies}

The SPM postulates the existence of a scalar potential $\rho$ (erg) field with the characteristics to cause the observed differences in spiral and elliptical galaxies.  The gradient of $\rho$ is proportional to a force $\vec{F}_\mathrm{s}$ (dyne) that acts on matter, 
\begin{equation}
\vec{F}_\mathrm{s} = G_\mathrm{s} m_\mathrm{s} \vec{\nabla} \rho
\label{eq:Fs},
\end{equation}
where the arrow over a parameter denotes a vector, the $G_\mathrm{s}$ is a proportionality constant analogous to the gravitational constant $G$, and $m_\mathrm{s}$ is a property of the test particle on which the $\vec{F}_\mathrm{s}$ acts.  Because the $m_\mathrm{s}$ property of matter is currently unidentified, call the units of $m_\mathrm{s}$ ``cs''.

The SPM suggests $\vec{F}_\mathrm{s}$ exerts a force to repel matter from spiral galaxies and to attract matter to early type galaxies.  Therefore, the $\rho$ is highest at the center of spiral galaxies as a point Source and lowest at the center of early type galaxies as a point Sink.  Call spiral galaxies Sources and early type galaxies Sinks.  

Because the scalar field $\rho_\epsilon$ due to Sources is highest at point Sources, it obeys the inverse square law as does gravity.  Therefore, 
\begin{equation}
\rho_\epsilon = K_\epsilon \epsilon /r_\epsilon
\label{eq:rsource},
\end{equation}
where $\epsilon$ is a number representing the strength of the Source analogous to the central mass of the gravitational potential, $ K_\epsilon $ is the proportionality constant, and $r_\epsilon$ (Mpc) is the distance from a Source to the point at which $\rho$ is calculated.  Because the nature of $\epsilon$ is currently unidentified, call the units of $\epsilon$ ``q''.  The unit of $ K_\epsilon $ is erg~Mpc~q$^{-1}$. 

Because the scalar field $\rho_\eta$ due to Sinks is lowest at point Sinks, it obeys the inverse square law as does gravity.  Therefore, 
\begin{equation}
\rho_\eta = K_\eta \eta /r_\eta
\label{eq:rsink},
\end{equation}
where $\eta$ (q) is a number representing the strength of the Sink analogous to the central mass of the gravitational potential, $ K_\eta $ (erg~Mpc~q$^{-1}$) is the proportionality constant, and $r_\eta$ (Mpc) is the distance from a Sink to the point at which $\rho$ is calculated.  

The $\rho$ is the sum of the effects of all galaxies,
\begin{equation}
\rho_\mathrm{p}  =  K_\epsilon \sum_{i=1}^{N_\mathrm{source}} \frac{\epsilon_i}{r_{\mathrm{p}i}} + K_\eta  \sum_{l=1}^{N_\mathrm{sink}} \frac{\eta_l}{r_{\mathrm{p}l}}
\label{eq:13b},
\end{equation}
where the subscript, lower case, italic, roman letters are indices; ``p'' is the point where $\rho_\mathrm{p}$ is evaluated; $r_{\mathrm{p}i}$ (Mpc) and $r_{\mathrm{p}l}$ (Mpc) are the distances to the point where $\rho_\mathrm{p}$ is evaluated from the $i^{th}$ Source and $l^{th}$ Sink, respectively; and $N_\mathrm{source}$ and $N_\mathrm{sink}$ are the number of Source and Sinks, respectively.

In the universe, $N_\mathrm{source}$ and $N_\mathrm{sink}$ are very large, perhaps effectively infinite.  The boundary condition of a very large universe produces considerable ambiguity in Eq.~(\ref{eq:13b}).  One way to resolve this ``Olber's paradox'' type condition is to postulate that $\eta <0$ and $\epsilon > 0$.  Call the volume with a simply connected surface containing equal Source and Sink strengths a ``cell''.  This is analogous to the electric charges in an atom.  As the universe may be considered a collection of clusters, the SPM considers the universe to be a collection of cells. 

Ignoring the effect of neighboring Sources and Sinks (the environment) and the radial movement of a test particle, the tangential rotation velocity $v_\theta$ (km~s$^{-1}$) of a test particle in a galaxy becomes
\begin{equation}
v_\theta^2 = \frac{G M_\mathrm{eff} }{r} 
\label{eq:49},
\end{equation}
where in a Source galaxy,
\begin{equation}
M_\mathrm{eff} = M - K_\epsilon \frac{G_\mathrm{s}}{G} \frac{m_\mathrm{s}}{m_\mathrm{g}} \epsilon 
\label{eq:52}, 
\end{equation}
in a Sink galaxy,
\begin{equation}
M_\mathrm{eff} = M + K_\eta \frac{G_\mathrm{s}}{G} \frac{m_\mathrm{s}}{m_\mathrm{g}} \vert \eta \vert 
\label{eq:52a}, 
\end{equation}
where, for simplicity, the mass of the test particle is assumed to be constant over time, $r$ (kpc) is the radial distance of the test particle, $M$ (M$_\odot$) is the mass inside the sphere of radius $r$, $\vert \, \vert$ indicates absolute value, and the inertial mass $m_\mathrm{\iota}$ equals gravitational mass $m_\mathrm{g}$ of the test particle \citep{will}.  

Because the outer RCs of elliptical galaxies are Keplerian \citep{roma}, the total force on mass must be centrally directed acting along the radius of the elliptical galaxy.  The $\vec{F}_\mathrm{s}$ and the gravitational force $\vec{F}_\mathrm{g}$ of the surrounding mass of the galaxy are directed radially inward.  Thus, the $M_\mathrm{eff}$ of a Sink galaxy appears greater than Newtonian dynamical expectation as the ``Virgocentric infall'' and ``Great Attractor'' phenomena require.  

In spiral galaxies the various rotation velocity curves can result if $\vec{F}_\mathrm{s}$ is directed radially outward and $\vec{F}_\mathrm{g}$ is directed radially inward with a $m_\mathrm{s}/m_\mathrm{g}$ ratio varing with $r$.  

The $\epsilon$ and $\eta$ terms of Eqs.~(\ref{eq:52}) and (\ref{eq:52a}) mimic a hidden mass.  However, because $M$ is the total mass, the $\epsilon$ and $\eta$ terms are massless.  

Because the H{\scriptsize{I}} and H$_\mathrm{\alpha}$ RCs differ, the $m_\mathrm{s}/m_\mathrm{g}$ factor must be different for different matter (elemental) types.  The $m_\mathrm{s}$ cannot be linearly related to the mass.  Another characteristic of matter must be the characteristic of $m_\mathrm{s}$.  Thus, the $M$ of a Source galaxy is greater than Newtonian dynamical expectation.  

The nature of the $\epsilon$ may (1) be derived from a property of matter, (2) be independent of matter, or (3) be the cause both of $\rho$ and of matter.  Because the $\rho$ field must be repulsive of matter in Source galaxies and attractive of matter in Sink galaxies, the first possibility is unlikely.  If matter falls into a galaxy before a Source is present, the galaxy is a Sink.  If a $\rho$ field of a Source exists before matter forms around it, the $\rho$ field prevents matter from forming a galaxy.  Hence the second possibility is unlikely.  Because Sources exist only with spiral galaxies around them, the $\rho$ field and the $M$ of a spiral galaxy are formed concurrently.  Therefore, $\epsilon \propto M_{\mathrm{t} \epsilon}$, where $ M_{\mathrm{t} \epsilon}$ is the total mass of the Source galaxy.  Similarly, $\eta \propto M_{\mathrm{t} \eta}$, where $ M_{\mathrm{t} \eta}$ is the total mass of the Sink galaxy.

Because Sinks acquire their mass from neighboring Source galaxies, if the Sink is in a low galaxy population density environment, then local Source galaxies yield little matter.  Such Sinks have little matter, have low luminosity, and have low $ \eta $.  Because of the low density of galaxies, the $ F_\mathrm{s}$ of neighboring galaxies is insignificant.  The extent of the $ F_\mathrm{s}$ field is limited to nearly the extent of the galaxy mass.  Because $ M_\mathrm{eff} > M$, measurements of lensing \citep{guzi} and velocities of satellites outside the Sink galaxy measure more mass using Newtonian dynamics than $M_{\mathrm{t} \eta}$ \citep{prad}.  This is a good description of $L^*$ elliptical galaxies \citep{roma}.

If the Sink is in a dense environment, then the Sink has more matter, is bright, and has high $\eta $.  The higher values of $F_\mathrm{s}$ and $F_\mathrm{g}$ compresses the matter (stars) well inside the extent of the $ F_\mathrm{s}$ field and the mass profile is steeper than low luminosity Sinks.  Because of the high density of galaxies, the $ F_\mathrm{s}$ of neighboring galaxies is significant.  Thus, the $ F_\mathrm{s}$ field is spatially more extended than the mass.  Data from lensing, stellar dynamics \citep{treu}, and velocities of satellites \citep{prad} outside the mass concentration of the Sink indicates much more mass than $M_{\mathrm{t} \eta}$ and the presence of a mass-like component more extended than the luminous component using Newtonian dynamics.  This is a good description of bright elliptical galaxies.

These observations are consistent with the premise that irregular, elliptical, and lenticular galaxies are Sinks and galaxies with spiral morphology are Sources.

Because Spiral galaxies form around Sources and because cD galaxies are often found in the center of large clusters with elliptical galaxies, cD galaxies were also considered Sinks.

\section{\label{sec:model}Redshift model}

The SPM suggests the photon is a particle.  The derivation of Planck's black body radiation equation includes the proposition that the energy of a photon is discrete and composed of a number $N$ of basic energy/particle packets.  The calculated redshift $z_\mathrm{c}$ is
\begin{equation}
z_\mathrm{c} \equiv \frac{\triangle \lambda}{\lambda} = \frac{N_\mathrm{e}}{N_\mathrm{o}}-1
\label{eq:3},
\end{equation}
where $\lambda$ is the wavelength, $N_\mathrm{e}$ is the emitted $N$, and $N_\mathrm{o}$ is the observed $N$.

Because the energy of the photon is changed as it traverses the $\rho$ field, $N$ is changed by the amount of $\rho$ through which the photon travels.  The change of $N$ per unit volume $V$ of $\rho$ through which the photon travels is posited to be proportional to $N $ entering $V$,
\begin{equation}
\frac{\mathrm{d} N}{\mathrm{d} V} = -K_\mathrm{v} N
\label{eq:4},
\end{equation}
where $K_\mathrm{v}$ is the proportionality constant.

The $V$ is
\begin{equation}
V= C_\mathrm{s} \, D \, \langle{\rho} \rangle 
\label{eq:7},
\end{equation}
where $\langle{\rho} \rangle$ is the average $\rho$ in $V$ and $C_\mathrm{s}$ is the cross section of $V$ which is the cross section of the photon particle.

Posit that a photon has a minimum number $N_\mathrm{min}$ of energy packets.  Combining Eqs.~(\ref{eq:3}) and (\ref{eq:4}) yields,
\begin{equation}
z_\mathrm{c} +1= \frac{N_\mathrm{e}}{N_\mathrm{min} + N_\mathrm{e} \, \exp({-K_\mathrm{v} \, \int_0^D \! \mathrm{d}V})}
\label{eq:6}.
\end{equation}

For the $D$ and change of $N$ considered herein, $C_\mathrm{s}$ is considered a constant.  For greater distances, where the total change of $N$ is relatively larger than considered herein, the $C_\mathrm{s}$ is a function of $N$ and $\rho$ at the position of the photon.  

The 
\begin{equation}
\langle{\rho} \rangle = \frac{1}{D} \, \int_0^D \rho_\mathrm{x} \, \mathrm{d}x
\label{eq:8},
\end{equation}
where d$x$ is the incremental distance $x$ (Mpc) traveled by the photon.  

The emitted luminosity of a galaxy is proportional to the flux of photons from the galaxy and is assumed to be isotropic.  Other factors such as the K--correction were considered too small for the $D$ of the sample.  For a spiral galaxy to exist, $F_\mathrm{g}$ must balance the $F_\mathrm{s}$.  Matter is ejected until this balance is achieved.  Therefore, the mass and, hence, the luminosity of a Source galaxy is a first order approximation of $\epsilon$.  Therefore, 
\begin{equation}
\epsilon = K_ {\epsilon \mathrm{l}} \, 10^{-0.4 \, M_{\beta }} 
\label{eq:11},
\end{equation}
where
\begin{equation}
\frac{ M_{\beta }}{\mathrm{mag.}} = \frac{m_{\beta }}{\mathrm{mag.}} - \frac{E_\mathrm{xt}}{\mathrm{mag.}} +25 - 5 \, \log_{10} \left(\frac{D}{\mathrm{Mpc}} \right)
\label{eq:12a};
\end{equation}
$ K_ {\epsilon \mathrm{l}} $ (q~s~erg$^{-1}$) is a proportionality constant; and $M_{\beta }$ (mag.), $m_{\beta }$ (mag.), and $E_{\mathrm{xt} }$ (mag.) are the absolute magnitude, apparent magnitude, and extinction, respectively, in a given band ($\beta$) of the galaxy. 

Similarily, for Sink galaxies,
\begin{equation}
\eta = K_{\eta \mathrm{l}}\, 10^{-0.4 \, M_{\beta }} 
\label{eq:11a},
\end{equation}
where $ K_{\eta \mathrm{l}}$ (q~s~erg$^{-1}$) is a proportionality constant.

The conservation of matter/energy implies for each Source galaxy
\begin{equation}
K_\mathrm{c} L_{\mathrm{I}} + M_{\mathrm{I}} = K_\mathrm{c} \epsilon + M_{\mathrm{O}} + \Delta M_{\mathrm{g}}
\label{eq:9},
\end{equation}
where $M_{\mathrm{I}}$ ($M_\odot$~s$^{-1}$), $M_{\mathrm{O}}$ ($M_\odot$~s$^{-1}$), $\Delta M_{\mathrm{g}}$ ($M_\odot$~s$^{-1}$), $L_{\mathrm{I}}$ (erg~s$^{-1}$), and $K_\mathrm{c}$ ($M_\odot$/erg) are the matter per second (exclusive of photons) entering the galaxy from other galaxies, the matter per second (exclusive of photons) emitted by the galaxy, the matter per second increase in the galaxy, the luminosity due to photons into the galaxy, and the proportionality constant to convert $\epsilon$ units to $M_\odot$~s$^{-1}$, respectively.  Matter is ``in'' a galaxy when it is in orbit around the Source, where the dynamics of the matter is principally determined by the galaxy parameters.

Similarly for Sink galaxies,
\begin{equation}
K_\mathrm{c} L_{\mathrm{I}} + M_{\mathrm{I}} = K_\mathrm{c} \eta + M_{\mathrm{O}} + \Delta M_{\mathrm{g}}
\label{eq:9a}.
\end{equation}

Sum Eqs.~(\ref{eq:9}) and (\ref{eq:9a}) over all Source and Sink galaxies, respectively.  Because the mass ejected from Source galaxies go to Sink galaxies, the combined $\Delta M_{\mathrm{g}} = 0$.  Because the measured CMB radiation is nearly an ideal black body radiation \citep{math, math2}, the sum of the $L_\mathrm{I}$ terms for all Sources equals the sum of the $L_\mathrm{I}$ terms for all Sinks.  Equating the sum of all matter emitted from Source galaxies to the sum of all matter entering Sink galaxies yields,  
\begin{equation}
\frac{ K_\eta}{ K_\epsilon} =\, -\, \frac{\sum_{i=1}^{N_\mathrm{sources}} \epsilon_i }{\sum_{k=1}^{N_\mathrm{sink}} \eta_k } 
\label{eq:28a},
\end{equation}
where $ K_{\eta \mathrm{l}} / K_{\epsilon \mathrm{l}} = { K_\eta}/{ K_\epsilon}$.

Because $V$ is a function of $x$ from the observer to a point along the line of sight and time $t$,
\begin{equation}
\mathrm{d}V(x,t)=\frac{\partial V}{\partial x}\mathrm{d} x + \frac{\partial V}{\partial t}\mathrm{d} t
\label{eq:14}.
\end{equation}

Combining Eqs.~(\ref{eq:7}), (\ref{eq:6}), (\ref{eq:8}), and (\ref{eq:14}) yields,
\begin{equation}
\frac{1}{z_\mathrm{c}+1}= K_\mathrm{min} + \mathrm{e}^X
\label{eq:26},
\end{equation}
where
\begin{equation}
X= K_\mathrm{dp} D P + K_\mathrm{d} D + K_\mathrm{p} P + K_\mathrm{f} F + K_\mathrm{vp} P v_\mathrm{e}
\label{eq:27}
\end{equation}
where: (1) Relatively small terms such as terms involving the relative $\epsilon$ of the emitter and observer were ignored.
(2)  The $ K_\mathrm{min} =N_\mathrm{min}/N_\mathrm{e}$ is a constant for a given $N_\mathrm{e}$.
(3)  The $P= \int_0^D \! \rho_\mathrm{p} \mathrm{d} x$.
(4)  The $F = \int_0^D \! [(\partial \rho_\mathrm{p} / \partial x)-K_\mathrm{co}] \mathrm{d} x$ to a first order approximation.
(5)  The $K_\mathrm{dp}$ (erg$^{-1}$~Mpc$^{-2}$), $K_\mathrm{d}$ (Mpc$^{-1}$), $K_\mathrm{p}$ (erg$^{-1}$~Mpc$^{-1}$), $K_\mathrm{f}$ (erg$^{-1}$), $K_\mathrm{co}$ (erg~Mpc$^{-1}$), and $K_\mathrm{vp}$ (erg$^{-1}$~Mpc$^{-1}$~deg.$^{-1}$)are constants.
(6)  The $\rho$ is posited to be constant over time.
(7)  The relative velocity of the emitting and observing galaxies causes a change in $V$, hence $N$, and has three possible causes.  One is the expansion of our universe.  This component is linearly related to $(K_\mathrm{dp} P + K_\mathrm{d}) D$.  The second cause is due to the possible peculiar velocity of the Milky Way relative to the reference frame derived by summing over all Sources and Sinks.  Another cause derives from the inaccuracy of defining the reference frame because the galaxies directly on the other side of the Milky Way center from earth are unobservable from earth.  The component $v_\mathrm{e}$ deriving from the second and third causes is proportional to the cosine of the angular difference between the direction of the target galaxy and the direction of $v_\mathrm{e}$.  Thus,
\begin{eqnarray}
v_\mathrm{e} &=& \cos(90^\circ - G_\mathrm{lat}) \, \cos(90^\circ -K_\mathrm{lat}) \, \nonumber \\*
& &+ \, \sin(90^\circ - G_\mathrm{lat}) \, \sin(90^\circ -K_\mathrm{lat}) \, \nonumber \\
& & \times \, \cos(G_\mathrm{lon}-K_\mathrm{lon})
\label{eq:16},
\end{eqnarray}
where $G_\mathrm{lat}$ (degrees) and $G_\mathrm{lon}$ (degrees) are the galactic latitude and longitude, respectively, of the emitting galaxy; and $K_\mathrm{lat}$ (degrees) and $K_\mathrm{lon}$ (degrees) are the galactic latitude and galactic longitude, respectively, of the direction of $v_\mathrm{e}$.

Setting $K_\epsilon=1$ considers $K_\epsilon $ was incorporated into the constants of Eq.~(\ref{eq:27}) and yields, 
\begin{equation}
K_\mathrm{\eta } = - \frac{\sum_{i=1}^{N_\mathrm{sources}} \epsilon_i }{\sum_{k=1}^{N_\mathrm{sink}} \eta_k } \, \, \mathrm{erg\,Mpc\,q}^{-1}
\label{eq:28}.
\end{equation}

\section{\label{sec:results}Results}

The sample galaxies were selected from the NED database\footnote{The Ned database is available at http://nedwww.ipac.caltech.edu.  The data were obtained from NED on 5 May 2004.}.  The selection criteria were that the heliocentric redshift $z_\mathrm{mh}$ be less than 0.03 and that the object be a galaxy.  The parameters obtained from the NED database included the galaxy name, $G_\mathrm{lon}$, $G_\mathrm{lat}$, $z_\mathrm{mh}$, morphology, the $m_{\mathrm{\beta}}$ was the B-band apparent magnitude $m_\mathrm{b}$ (mag.) as defined by NED, and the galactic extinction $E_{\mathrm{xt}}$ (mag.).  The $z_{\mathrm{m}}$ was calculated from $z_\mathrm{mh}$. 

The 21-cm line width $W_\mathrm{20}$ (km~s$^{-1}$) at 20 percent of the peak and the inclination $i_\mathrm{n}$ (arcdegrees) between the line of sight and polar axis were obtained from the LEDA database\footnote{The LEDA database is available at http://leda.univ-lyon.fr.  The data were obtained from LEDA on 5 May 2004.} when such data existed.

The constants to be discovered are the constants of Eq.~(\ref{eq:26}), the cell limitation of $r_{\mathrm{p}i}$ and $r_{\mathrm{p}l}$, and $K_\eta$.  Calculating the constants was done by making the following simplifications: (1) Estimate the $D$ to the sample galaxies (see Appendix~\ref{sec:initial}).  (2) Galaxies with an unlisted morphology in the NED database were considered to have negligible effect.  An alternate method may be to assign a high value of $m_{\mathrm{b}}$ to such galaxies.  This option was rejected because a large number of such galaxies were from the 2dFGRS and 2MASS.  These surveys include only limited areas of the sky.  Therefore, including the 2dFGRS and 2MASS galaxies with unlisted morphology in the calculation would introduce a selection bias into the sample.  (3) Galaxies with an unlisted $m_\mathrm{b}$ were assigned $M_\mathrm{b} = -11$ mag.  (4) Objects with $E_{\mathrm{xt}} = 99$ in NED were assigned an $E_{\mathrm{xt}}=0$ mag.  (5) All the sample galaxies were considered mature and stable.  (6) Galaxies with a spiral (non-lenticular), barred, or ringed morphology in the NED database were considered Sources.  All other galaxies were considered Sinks.  The result was a sample of 22,631 Source galaxies and 7,268 Sink galaxies. 

Define the luminosity ratio $L_\mathrm{r}$ as the ratio of the sum of the luminosities of all Sources within a limiting distance $D_\mathrm{l}$ divided by the sum of the luminosities of all Sinks within $D_\mathrm{l}$,
\begin{equation}
L_\mathrm{r} = \frac{\sum_{i=1}^{N_\mathrm{sources}(D< D_\mathrm{l})} \epsilon_i }{\sum_{k=1}^{N_\mathrm{sink}(D< D_\mathrm{l})} \eta_k } 
\label{eq:28z}.
\end{equation}

\begin{figure}
\begin{center}
\includegraphics[width=0.8\textwidth]{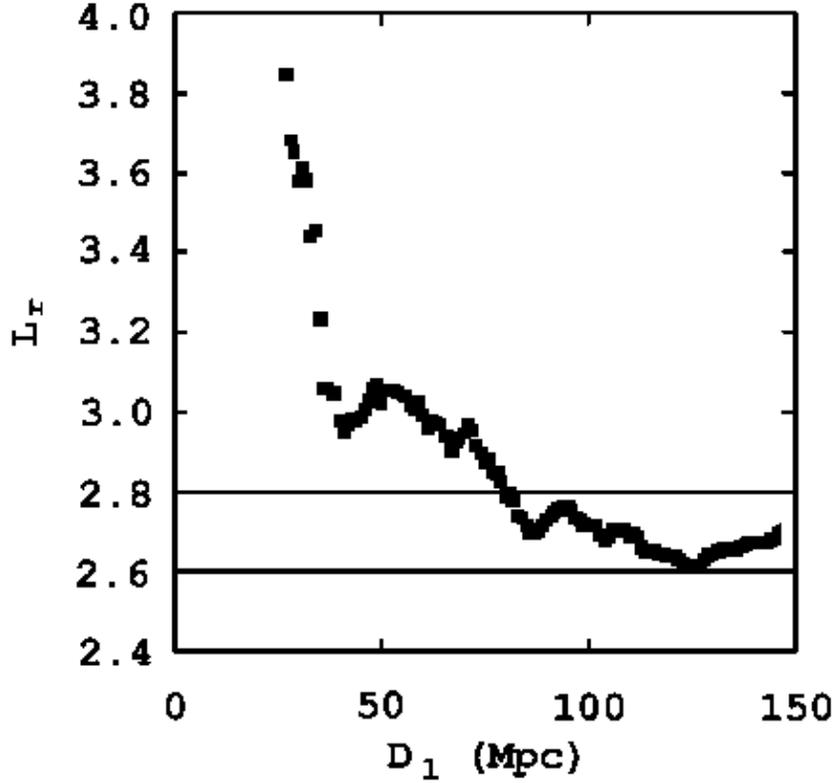}
\end{center}
\caption{Plot of the luminosity ratio $L_\mathrm{r}$ for galaxies with a distance $D$ less than the limiting distance $D_\mathrm{l}$ versus $D_\mathrm{l}$.}
\label{fig:Lum}
\end{figure}

Figure~\ref{fig:Lum} shows a plot of $L_\mathrm{r}$ versus $D_\mathrm{l}$ for the 29,899 sample galaxies.  Because the sample galaxies were limited to $z_\mathrm{mh}< 0.03$, the selection of galaxies for $D > 130$~Mpc was incomplete.  Therefore, from Eq.~(\ref{eq:28}) \mbox{$ K_\mathrm{\eta } = -2.7 \pm 0.1$~erg~Mpc~q$^{-1}$}.

Form the linear relation, 
\begin{equation}
z_\mathrm{c} = K_\mathrm{scm} z_\mathrm{m} + K_\mathrm{icm}
\label{eq:35},
\end{equation}
where $K_\mathrm{scm}$ is the least squares slope and $K_\mathrm{icm}$ is the least squares intercept of the presumed linear relationship between $z_\mathrm{c}$ and $ z_\mathrm{m}$.  The constants of Eq.~(\ref{eq:26}) were adjusted to maximize the correlation coefficient of Eq.~(\ref{eq:35}) with $K_\mathrm{scm} \approx 1$ and with $ K_\mathrm{icm} \approx 0$.  

\begin{figure}
\begin{center}
\includegraphics[width=0.8\textwidth]{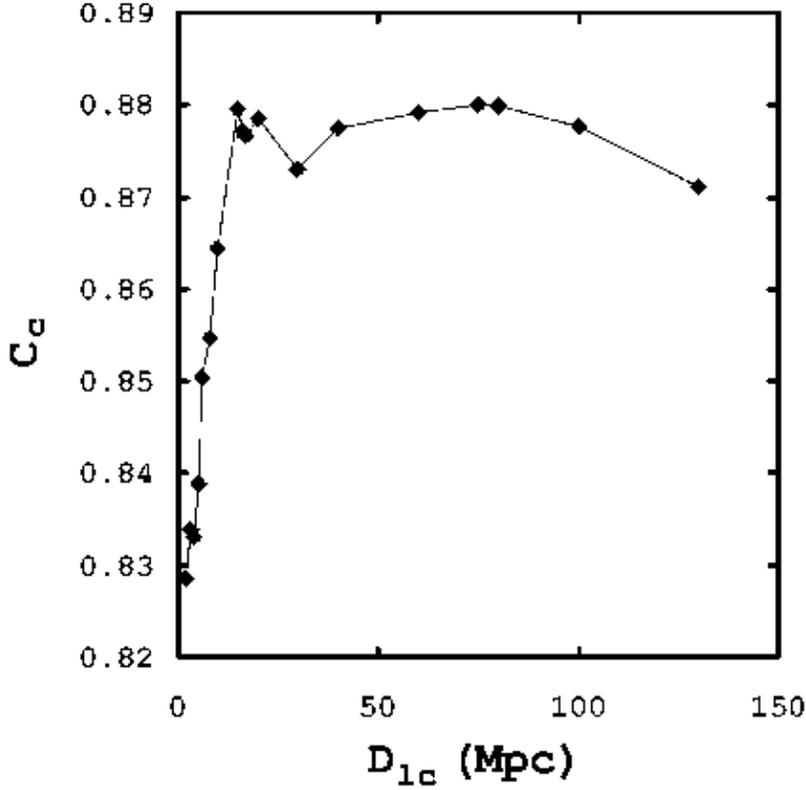}
\end{center}
\caption{Plot of the correlation coefficient $C_\mathrm{c}$ of Eq.~(\ref{eq:35}) using the best values of the constants to be discovered versus the distance $D_\mathrm{lc}$ (Mpc) limitation of $r_{\mathrm{x}i}$ and $r_{\mathrm{x}l}$. }
\label{fig:CcvsDlc}
\end{figure}

Figure~\ref{fig:CcvsDlc} shows a plot of the correlation coefficient $C_\mathrm{c}$ of Eq.~(\ref{eq:35}) using the best values of the constants to be discovered for each data point versus the distance $D_\mathrm{lc}$ (Mpc) limitation of $r_{\mathrm{p}i}$ and $r_{\mathrm{p}l}$.  Peaks of $C_\mathrm{c} \approx 0.88$ were obtained at $D_\mathrm{lc} = 15$~Mpc and $D_\mathrm{lc}=75$~Mpc.  Therefore, the $r_{\mathrm{p}i}$ and $r_{\mathrm{p}l}$ were limited to 15~Mpc.  Of the sample galaxies, 3,480 Source galaxies and 1,604 Sink galaxies were within 15~Mpc of at least one of the Category A galaxies.  Note the distances to the close, major clusters (Virgo and Fornax) vary between 15 Mpc and 20 Mpc.  The distance of the next farther clusters (Pisces, Perseus, and Coma) vary between 40~Mpc and 60~Mpc.  

\begin{figure}
\begin{center}
\includegraphics[width=0.8\textwidth]{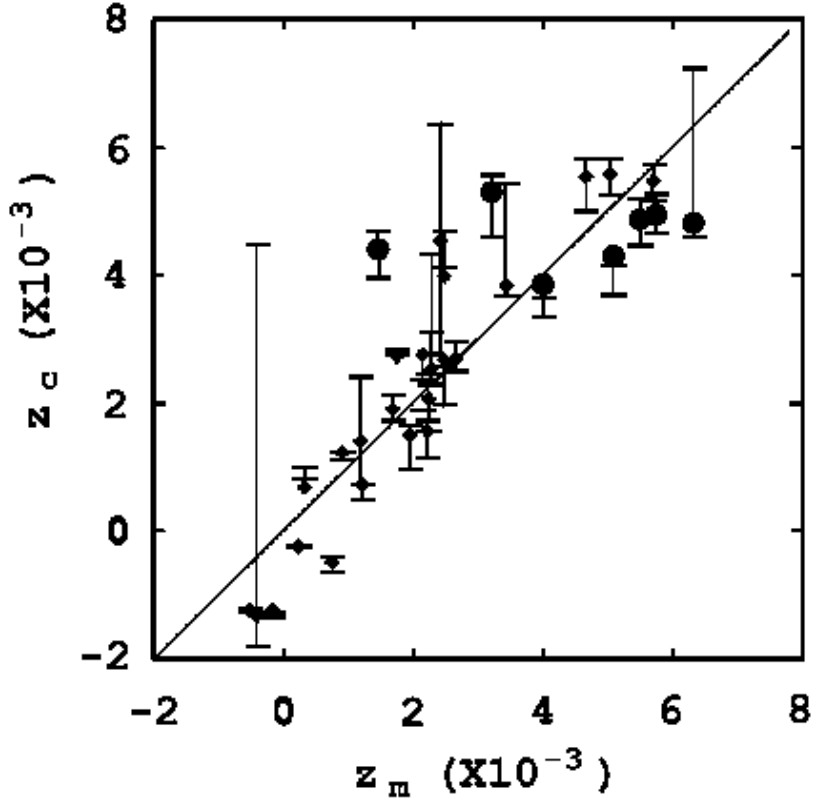}
\end{center}
\caption{Plot of the calculated redshift $z_\mathrm{c}$ versus the measured redshift $z_\mathrm{m}$ for 32 Category A galaxies \citep{free,macr}.  The straight line is a plot of $z_\mathrm{c} = z_\mathrm{m}$.  The circles indicate the data points for galaxies with (l,b) = (290$^\circ \pm 20^\circ$,75$^\circ \pm 15^\circ$).}
\label{fig:7}
\end{figure}

Figure~\ref{fig:7} shows a plot of $z_\mathrm{c}$ versus $z_{\mathrm{m}}$ for the 32 Category A galaxies.  Tables~\ref{tab:1} and \ref{tab:2} lists the data for the 32 Category A galaxies.  Table~\ref{tab:4} lists the calculated constants of Eq.~(\ref{eq:26}).  The $K_\mathrm{scm} = 1.0 \pm 0.1$ and $K_\mathrm{icm}= (0 \pm 9) \times 10^{-5} $ at 1$\sigma$ with a correlation coefficient of 0.88. 

The error bars indicate the $z_\mathrm{c}$ for 1.05$D_\mathrm{a}$ and 0.95$D_\mathrm{a}$, which is consistent with the error cited in \citet{free}.  For some galaxies such as \astrobj{NGC~4548} both of the recalculations yield a higher $z_\mathrm{c}$ calculation than the calculation using $D_\mathrm{a}$.  
%\begingroup
%\squeezetable
\begin{table*}
\scriptsize
\caption{Category A galaxy data in order of increasing $D_\mathrm{a}$.}
\label{tab:1}
\begin{tabular}{llrrrrrrrrr}
\hline
\hline
{Galaxy}
&{Morphology}$^\mathrm{a}$
&$G_\mathrm{lon}$
&{$G_\mathrm{lat}$}
&{$D_\mathrm{a}$} 
&{$z_\mathrm{m}$}
&{$z_\mathrm{h}$}
&{$z_\mathrm{c}$ }
&{$P$}
&{$F$}\\
& &&&& $\times 10^{-3}$&$\times 10^{-3}$ &$\times 10^{-3}$ & $\times 10^{10}$& $\times 10^{12}$ \\
& &deg.&deg.&Mpc&& & &erg~Mpc&erg\\
\hline
\astrobj{IC 1613}&IB(s)m&130&-61&0.65&-0.518&0.156&-1.234&0.25&0.04\\
\astrobj{NGC~0224}&SA(s)b LINER&121&-22&0.79&-0.408&0.190&-1.327&0.34&0.02\\
\astrobj{NGC~0598}&SA(s)cd HII&134&-31&0.84&-0.148&0.202&-1.240&0.32&0.04\\
\astrobj{NGC~0300}&SA(s)d&299&-79&2.00&0.336&0.480&0.687&0.81&0.09\\
\astrobj{NGC~5253}&Im pec;HII Sbrst&315&30&3.15&0.903&0.756&1.219&0.82&0.15\\
\astrobj{NGC~2403}&SAB(s)cd HII&151&29&3.22&0.758&0.773&-0.479&0.80&0.15\\
\astrobj{NGC~3031}&SA(s)ab;LINER Sy1.8&142&41&3.63&0.243&0.871&-0.249&0.92&0.15\\
\astrobj{IC 4182}&SA(s)m  &108&79&4.49&1.231&1.078&0.735&1.01&0.21\\
\astrobj{NGC~3621}&SA(s)d  &281&26&6.64&1.759&1.594&2.744&1.37&0.29\\
\astrobj{NGC~5457}&SAB(rs)cd  &102&60&6.70&1.202&1.608&1.406&2.20&0.07\\
\astrobj{NGC~4258}&SAB(s)bc;LINE  R Sy1.9 &138&69&7.98&1.694&1.915&1.896&1.81&0.34\\
\astrobj{NGC~0925}&SAB(s)d  HII     &145&-25&9.16&2.216&2.198&1.566&1.68&0.40\\
\astrobj{NGC~3351}&SB(r)b;HII  Sbrst   &234&56&10.00&2.258&2.400&2.522&1.77&0.44\\
\astrobj{NGC~3627}&SAB(s)b;LINER  Sy2     &242&64&10.05&2.145&2.412&2.753&1.82&0.43\\
\astrobj{NGC~3368}&SAB(rs)ab;Sy  LINER   &234&57&10.52&2.659&2.525&2.703&1.89&0.45\\
\astrobj{NGC~2541}&SA(s)cd  LINER   &170&33&11.22&1.963&2.693&1.492&1.91&0.49\\
\astrobj{NGC~2090}&SA (rs)b  &239&-27&11.75&2.490&2.820&3.985&2.16&0.51\\
\astrobj{NGC~4725}$^\mathrm{b}$&SAB(r)ab pec  Sy2     &295&88&12.36&4.026&2.966&3.848&2.32&0.52\\
\astrobj{NGC~3319}&SB(rs)cd  HII     &176&59&13.30&2.497&3.192&2.669&2.38&0.58\\
\astrobj{NGC~3198}&SB(rs)c  &171&55&13.80&2.281&3.312&2.546&2.44&0.60\\
\astrobj{NGC~2841}&SA(r)b ;LINER  Sy1     &167&44&14.07&2.249&3.377&2.070&2.49&0.58\\
\astrobj{NGC~7331}&SA(s)b  LINER   &94&-21&14.72&3.434&3.533&3.850&2.12&0.62\\
\astrobj{NGC~4496A}$^\mathrm{b}$&SB(rs)m  &291&66&14.86&5.505&3.566&4.883&2.30&0.65\\
\astrobj{NGC~4536}$^\mathrm{b}$&SAB(rs)bc      HII  &293&65&14.93&5.752&3.583&4.963&2.31&0.65\\
\astrobj{NGC~4321}$^\mathrm{b}$&SAB(s)bc;LINER HII  &271&77&15.21&5.087&3.650&4.306&2.29&0.65\\
\astrobj{NGC~4535}$^\mathrm{b}$&SAB(s)c        HII  &290&71&15.78&6.325&3.787&4.803&2.32&0.68\\
\astrobj{NGC~1326A}&SB(s)m  &239&-56&16.14&5.713&3.874&5.478&2.51&0.71\\
\astrobj{NGC~4548}$^\mathrm{b}$&SBb(rs);LINER  Sy      &286&77&16.22&1.476&3.893&4.408&2.18&0.69\\
\astrobj{NGC~4414}&SA(rs)c?  LINER   &175&83&17.70&2.416&4.248&4.557&2.94&0.75\\
\astrobj{NGC~1365}&(R')SBb(s)b    Sy1.8  &238&-55&17.95&5.049&4.308&5.589&2.57&0.77\\
\astrobj{NGC~1425}&SA(rs)b  &228&-53&21.88&4.666&5.251&5.542&2.52&0.94\\
\astrobj{NGC~4639}$^\mathrm{b}$&SAB(rs)bc  Sy1.8   &294&76&21.98&3.223&5.275&5.315&2.33&0.95\\
\hline
\end{tabular}

$^\mathrm{a}${Galaxy morphological type from the NED database.}
$^\mathrm{b}${The data points for galaxies with (l,b) = (290$^\circ \pm 20^\circ$,75$^\circ \pm 15^\circ$). }
\end{table*}
%\endgroup

%\begingroup
%\squeezetable
\begin{table*}
\scriptsize
\caption{The components of Eq.~(\ref{eq:26}) for each Category A galaxy. }
\label{tab:2}
\begin{tabular}{lrrrrr}
\hline
\hline
{Galaxy}
&{$ K_\mathrm{dp} D P $}
&{$ K_\mathrm{p} P $}
&{$ K_\mathrm{f} F $}
&{$ K_\mathrm{vp} P v_\mathrm{e}$}
&{$X$}\\ 
&$\times 10^{-3}$&$\times 10^{-3}$&$\times 10^{-3}$&$\times 10^{-3}$&$\times 10^{-3}$\\
\hline
\astrobj{IC 1613}&0.009&-0.563&-0.204&0.06&-0.70\\
\astrobj{NGC~0224}&0.014&-0.770&-0.103&0.25&-0.61\\
\astrobj{NGC~0598}&0.014&-0.727&-0.226&0.25&-0.69\\
\astrobj{NGC~0300}&0.086&-1.836&-0.492&-0.38&-2.63\\
\astrobj{NGC~5253}&0.137&-1.862&-0.780&-0.65&-3.16\\
\astrobj{NGC~2403}&0.136&-1.808&-0.760&0.98&-1.46\\
\astrobj{NGC~3031}&0.177&-2.091&-0.778&1.00&-1.69\\
\astrobj{IC 4182}&0.240&-2.292&-1.067&0.45&-2.67\\
\astrobj{NGC~3621}&0.484&-3.121&-1.523&-0.52&-4.68\\
\astrobj{NGC~5457}&0.783&-5.005&-0.380&1.26&-3.34\\
\astrobj{NGC~4258}&0.767&-4.117&-1.761&1.28&-3.83\\
\astrobj{NGC~0925}&0.814&-3.807&-2.097&1.58&-3.50\\
\astrobj{NGC~3351}&0.939&-4.020&-2.267&0.89&-4.46\\
\astrobj{NGC~3627}&0.969&-4.129&-2.254&0.72&-4.69\\
\astrobj{NGC~3368}&1.054&-4.292&-2.337&0.93&-4.64\\
\astrobj{NGC~2541}&1.135&-4.333&-2.560&2.33&-3.43\\
\astrobj{NGC~2090}&1.344&-4.899&-2.661&0.29&-5.92\\
\astrobj{NGC~4725}$^\mathrm{a}$&1.521&-5.269&-2.711&0.67&-5.78\\
\astrobj{NGC~3319}&1.681&-5.414&-3.021&2.15&-4.61\\
\astrobj{NGC~3198}&1.786&-5.542&-3.116&2.39&-4.48\\
\astrobj{NGC~2841}&1.860&-5.662&-3.008&2.80&-4.01\\
\astrobj{NGC~7331}&1.656&-4.818&-3.234&0.61&-5.79\\
\astrobj{NGC~4496A}$^\mathrm{a}$&1.812&-5.223&-3.376&-0.03&-6.82\\
\astrobj{NGC~4536}$^\mathrm{a}$&1.825&-5.235&-3.363&-0.12&-6.90\\
\astrobj{NGC~4321}$^\mathrm{a}$&1.848&-5.204&-3.395&0.51&-6.24\\
\astrobj{NGC~4535}$^\mathrm{a}$&1.943&-5.274&-3.524&0.12&-6.74\\
\astrobj{NGC~1326A}&2.143&-5.687&-3.679&-0.19&-7.41\\
\astrobj{NGC~4548}$^\mathrm{a}$&1.875&-4.952&-3.605&0.34&-6.34\\
\astrobj{NGC~4414}&2.763&-6.685&-3.923&1.35&-6.49\\
\astrobj{NGC~1365}&2.447&-5.839&-3.989&-0.14&-7.52\\
\astrobj{NGC~1425}&2.919&-5.715&-4.910&0.23&-7.47\\
\astrobj{NGC~4639}$^\mathrm{a}$&2.718&-5.297&-4.920&0.25&-7.25\\
\hline
\end{tabular}

$^\mathrm{a}${The data points for galaxies with (l,b) = (290$^\circ \pm 20^\circ$,75$^\circ \pm 15^\circ$).}

\end{table*}
%\endgroup

%\begingroup
%\squeezetable
\begin{table*}
\caption{The values of the constants of Eq.~(\ref{eq:26}). }
\label{tab:4}
\begin{tabular}{lll}
\hline
\hline
{Parameter}
&{value}
&units\\ 
\hline
$K_\mathrm{min}$& \phantom{-1}$1.94 \times 10^{-3}$&\\
$K_\mathrm{dp}$& \phantom{-1}$5.3 \times 10^{-15}$& erg$^{-1}$~Mpc$^{-2}$\\
$K_\mathrm{d}$$^\mathrm{a}$& \phantom{-1}$0.00$& Mpc$^{-1}$\\
$K_\mathrm{p}$& $-2.27 \times 10^{-13}$& erg$^{-1}$~Mpc$^{-1}$\\
$K_\mathrm{f}$& $\,-5.2 \times 10^{-15}$& erg$^{-1}$\\
$K_\mathrm{co}$& $\,-4.3 \times 10^{10}$& erg~Mpc$^{-1}$\\
$K_\mathrm{vp}$& \phantom{-1}$1.3 \times 10^{-13}$&erg$^{-1}$~Mpc$^{-1}$~deg.$^{-1}$\\
$K_\mathrm{lat}$& \phantom{1}$14$&$^\circ$\\
$K_\mathrm{lon}$& $164$&$^\circ$\\
\hline\\
\end{tabular}

$^\mathrm{a}$ {Because $K_\mathrm{d}=0$, it is ignored in the final equation.}

\end{table*}
%\endgroup

If the non-target (other) galaxy is far from the photon's path, it has little individual effect.  If the other galaxy is close to the photon's path, an error in its distance changes the distance to the photon's path by the tangent of a small angle.  Also, because $M_\mathrm{b}$ is calculated using $D$, the slight change in $r_{\mathrm{p}i}$ and $r_{\mathrm{p}l}$ is partially offset by the change in $M_\mathrm{b}$.  Therefore, the error in $D$ of other galaxies is negligible relative to the effect of a $D_\mathrm{a}$ error of the target galaxy.  

\begin{figure}[!ht]
\begin{center}
\includegraphics[width=0.8\textwidth]{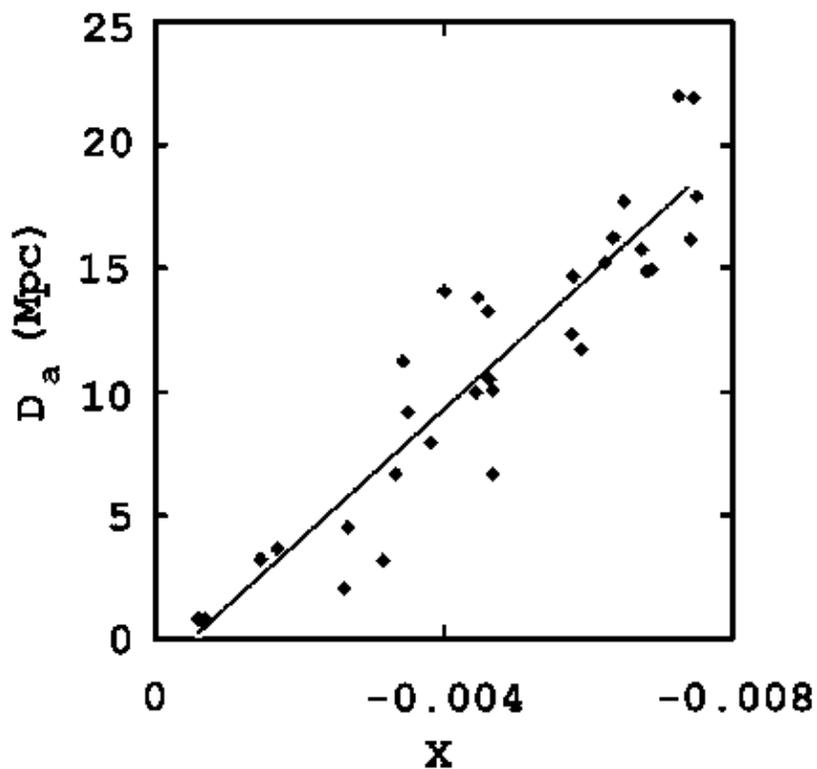}
\end{center}
\caption{Plot of distance $D_a$ (Mpc) versus exponent factor $X$ of the redshift calculation for 32 Category A sample galaxies.  The straight line indicates $D_a = -2600X - 1$.  }
\label{fig:2z}
\end{figure}

The Category A galaxies are within 22~Mpc from earth.  The $X$ term of Eq.~(\ref{eq:26}) predominates and $K_\mathrm{min}$ is relatively small for distances less than a few Gpc.  Therefore, $z \longrightarrow \exp( -X ) \,-1 \approx \, -X$.  Figure~\ref{fig:2z} is a plot of $D_\mathrm{a}$ versus $X$.  The straight line is a plot of the least squares fit of the data.  The line is 
\begin{eqnarray}
D_\mathrm{a}&=& (-2700 \pm 500 \mathrm{Mpc} ) X - (1.4 \pm 0.8 \mathrm{Mpc}) \nonumber \\*
&\approx& \frac{c}{H_\mathrm{spm}} z
\label{eq:28b}
\end{eqnarray}
at 1$\sigma$ and with a correlation coefficient of 0.93, where \mbox{$ H_\mathrm{spm} =110 \pm 20$~km~s$^{-1}$~Mpc$^{-1}$}.

At $D =18$ Gpc $\exp(X) \approx K_\mathrm{min} / 2$.  At large cosmological distance, \mbox{$z \longrightarrow \,K_\mathrm{min}^{-1} \approx 500$}.

\section{\label{sec:Xfactor}X factors} 

\begin{figure}[!ht]
\begin{center}
\includegraphics[width=0.8\textwidth]{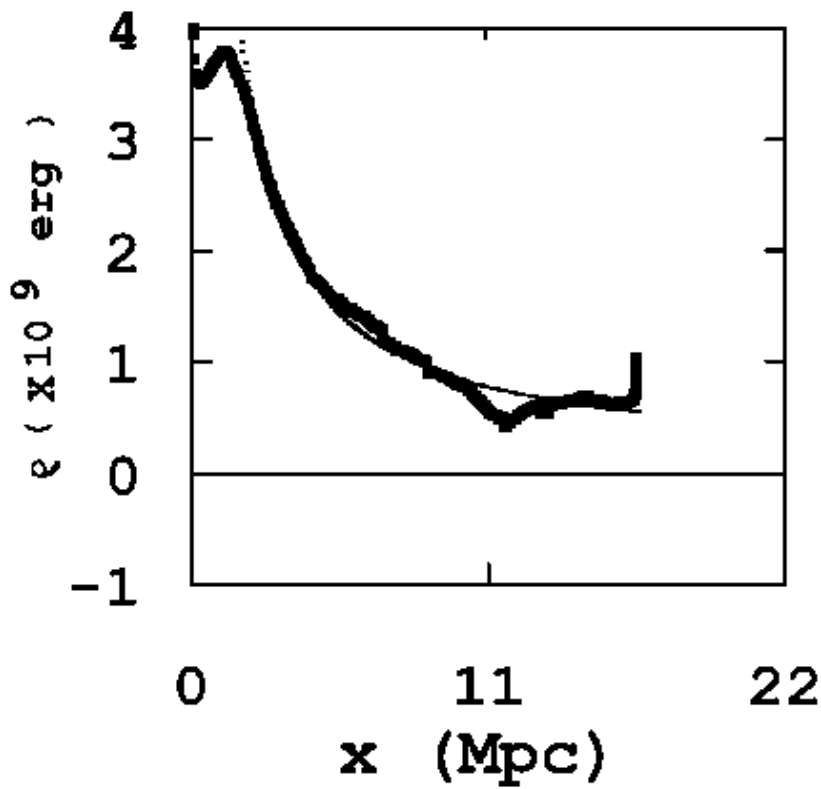}
\end{center}
\caption{Plot of scalar potential $\rho \, (\times 10^{9}$ erg) versus distance $x$ (Mpc) along our line of sight of \astrobj{NGC~1326A} $[ (l,b,z) \approx  (239^\circ, -56^\circ, 0.0057134) ]$.  }
\label{fig:6b}
\end{figure}

\begin{figure}[!t]
\begin{center}
\includegraphics[width=0.8\textwidth]{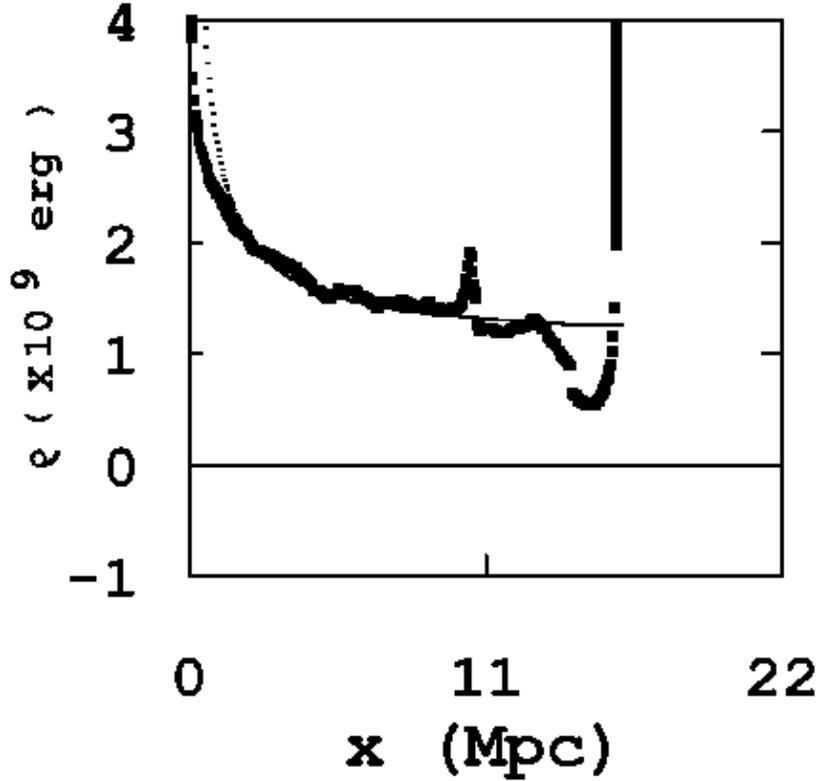}
\end{center}
\caption{Plot of scalar potential $\rho \, (\times 10^{9}$ erg) versus distance $x$ (Mpc) along our line of sight of \astrobj{NGC~4535} $[ (l,b,z) \approx  (290^\circ, 71^\circ, 0.0063253) ]$. }
\label{fig:6c}
\end{figure}

\begin{figure}[!ht]
\begin{center}
\includegraphics[width=0.8\textwidth]{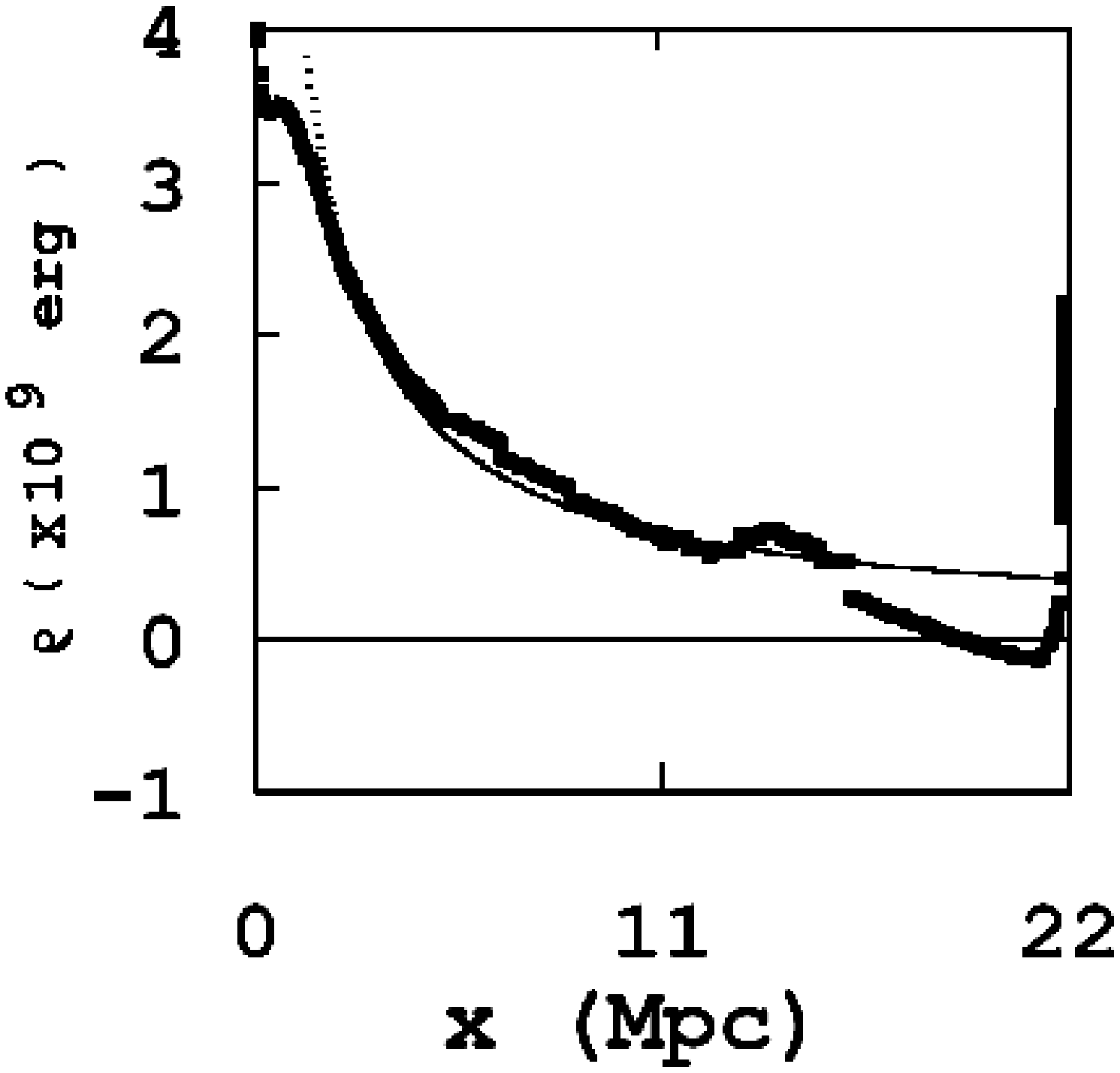}
\end{center}
\caption{Plot of scalar potential $\rho \, (\times 10^{9}$ erg) versus distance $x$ (Mpc) along our line of sight of \astrobj{NGC~1425} $[ (l,b,z) \approx  (228^\circ, -53^\circ, 0.0046664) ]$. }
\label{fig:6d}
\end{figure}

The components of $X$ in Eq.~(\ref{eq:27}) are a directional effect and effects calculated from the integration of $\rho_\mathrm{p}$and $\partial \rho_\mathrm{p} / \partial x $ across $\Delta x$.  The $X$ value may be calculated by the summation of the galaxy effects of the various regions along the light path.  The Milky Way effect $X_\mathrm{mw}$ is caused by the Source in the Galaxy.  Because it is close to earth, $\rho$ is high and limited to $x <0.05$~Mpc as shown in the examples \astrobj{NGC~1326A} (Fig.~\ref{fig:6b}), \astrobj{NGC~4535} (Fig.~\ref{fig:6c}), and \astrobj{NGC~1425} (Fig.~\ref{fig:6d}). 

The directional dependant effect $X_\mathrm{dir}$ is caused by the $v_\mathrm{e}$ term of Eq.~(\ref{eq:26}) and the Sources and Sinks in the local group, over 0.05~Mpc $ <x <2$~Mpc.  The variation in $X_\mathrm{dir}$ due to the cluster structure of our local group is seen in Figs.~\ref{fig:6b}, \ref{fig:6c}, and \ref{fig:6d}.  Figure~\ref{fig:6b} shows the light from the target galaxy passing first through the outer shell of Source galaxies of the local cluster with a large hump in the $\rho$ -- $x$ plot and then through the Sink galaxies of the core of the local cluster with a large ``U'' in the $\rho$ -- $x$ plot.  Figure~\ref{fig:6d} shows light from the target galaxy passing first through a portion of the outer shell of Source galaxies of the local cluster with a small hump in the $\rho$ -- $x$ plot and then near but not through the Sink galaxies of the core of the local cluster with a barely discernable ``U'' in the $\rho$ -- $x$ plot.  Figure~\ref{fig:6c} shows the light from the target galaxy passing through only a portion of the outer shell of Source galaxies of the local group with neither a hump nor a ``U'' being discernable.  

The $D^{-1}$ effect $X_\mathrm{d}$ at $x > 2$~Mpc is caused by the local group that appears as a single Source with a $\rho$ value depending on direction as expected for a Gaussian surface.  If there were no other influences, Eq.~(\ref{eq:rsource}) expects a $D^{-1}$ decline in $\rho$.  This is plotted in Figs.~\ref{fig:6b}, \ref{fig:6c}, and \ref{fig:6d} as dashes.  Noting $P \propto \int \rho \mathrm{d} x \approx \ln D$ and terms of Eq.~(\ref{eq:27}) other than $P$ terms are small if the only $\rho$ variation is $D^{-1}$, using only first order terms, and performing a binomial expansion yields $z$ is a linear function of $D$ (the Hubble Law) with a negative (blueshift) intercept, which is consistent with Eq.~(\ref{eq:28b}) and consistent with observations of close galaxies. 

As shown in Figs.~\ref{fig:6b}, \ref{fig:6c}, and \ref{fig:6d} the $\rho$ varies both above and below $X_\mathrm{d}$.  An $\int \rho \mathrm{d} x$ above $X_\mathrm{d}$ implies a $\mathrm{d} z /\mathrm{d} D$ greater than the $X_\mathrm{d}$ caused $z$ change.  An $\int \rho \mathrm{d} x$ below $X_\mathrm{d}$ implies a $\mathrm{d} z /\mathrm{d} D$ less than the $X_\mathrm{d}$ caused $z$ change.  

The field Source effect $X_\mathrm{source}$ is caused by the light passing very close to a Source galaxy outside of a cluster.  This effect is seen in the plot of \astrobj{NGC~4535} (Fig.~\ref{fig:6c}) as a sharp peak at $x \approx 10.4$~Mpc. The $X_\mathrm{source}$ in the plot of \astrobj{NGC~4535} (Fig.~\ref{fig:6c}) appears to be caused by \astrobj{NGC~4526} (SAB, $D_\mathrm{b}=10.41$~Mpc, $M_\mathrm{b} = -19.44$ mag., $A=0.5^\circ$), where $A$ (arcdegrees) is the angular separation of other galaxies from the identified target galaxy.  

The field Sink effect $X_\mathrm{sink}$ is caused by the light passing very close to a Sink galaxy outside of a cluster.  This effect is seen in the plot of \astrobj{NGC~4535} (Fig.~\ref{fig:6c}) as a sharp ``V'' at $x \approx 5.6$~Mpc. The $X_\mathrm{sink}$ in the plot of \astrobj{NGC~4535} (Fig.~\ref{fig:6c}) appears to be caused by \astrobj{Messier 095} (E5, $D_\mathrm{e}=5.59$~Mpc, $M_\mathrm{b} = -18.23$ mag., $A=3.9^\circ$).  

The galaxy group effect $X_\mathrm{g}$ is seen in the plots of \astrobj{NGC~1326A} (Fig.~\ref{fig:6b}) and \astrobj{NGC~1425} (Fig.~\ref{fig:6d}) as sharp steps associated with extended $\rho$ values slightly higher than $X_\mathrm{d}$ at $x \approx 7$~Mpc and $x \approx 8.5$~Mpc.  The step and extended areas of \astrobj{NGC~1326A} and \astrobj{NGC~1425} (Fig.~\ref{fig:6d}) are caused by several low luminosity Source galaxies intermixed with a few low luminosity Sink galaxies.  Unlike in clusters, the field galaxies have a higher percentage of Source galaxies and a lower density \citep{binn}.  The $X_\mathrm{g}$ is distinguished from $X_\mathrm{source}$ by the larger $\Delta x > 0.5$~Mpc of the group. 

The total cell effect $X_\mathrm{tc}$ is caused by light passing through or near the center of a cluster.  The $X_\mathrm{tc}$ is a broad ``V'' shape.  This is shown in \astrobj{NGC~1326A} (Fig.~\ref{fig:6b}) at $x= 11.6$~Mpc, \astrobj{NGC~4535} (Fig.~\ref{fig:6c}) at $x= 11.6$~Mpc, and \astrobj{NGC~1425} (Fig.~\ref{fig:6d}) at $x= 12.6$~Mpc.  Within 5.0$^\circ$ and 4~Mpc of the point of the ``V'', \astrobj{NGC~1326A} has 14 sinks and 1 source, \astrobj{NGC~4535} has 34 Sinks and 20 Sources, and \astrobj{NGC~1425}  has 5 Sinks and zero Sources. 

Another cluster effect $X_\mathrm{shell}$ is caused by the high density of Source galaxies in the shell of a cluster.  This is shown in \astrobj{NGC~1326A} (Fig.~\ref{fig:6b}) at $x= 14.4$~Mpc, \astrobj{NGC~4535} (Fig.~\ref{fig:6c}) at $x= 12.9$~Mpc, and \astrobj{NGC~1425} (Fig.~\ref{fig:6d}) at $x= 14.4$~Mpc.  

Another cluster effect $X_\mathrm{core-}$ is caused by the high density of Sinks galaxies in the core of a cluster and by the target galaxy being in the shell of a cluster.  This is shown in \astrobj{NGC~4535} (Fig.~\ref{fig:6c}) at $x= 15$~Mpc and \astrobj{NGC~1425} (Fig.~\ref{fig:6d}) at $x= 21.1$~Mpc.  This effect is caused by light passing through the core of a cluster.  Therefore, the core acts as a large Sink.  This effect occurs only if the target galaxy is in the far side of the cluster from us.

Another cluster effect $X_\mathrm{core+}$ is caused by the high density of Sources galaxies in the shell of a cluster and by the target galaxy being in the shell of a cluster.  This is shown in \astrobj{NGC~1326A} (Fig.~\ref{fig:6b}) at $x= 15.5$~Mpc.  This effect occurs only if the target galaxy is in the near side of the cluster from us.  Although there is a slight dip in the $\rho - D$ plot, the net effect is the $\rho$ values are higher than the $X_\mathrm{d}$ values.  

The net effect $X_\mathrm{core}$ is either the $X_\mathrm{core+}$ or $X_\mathrm{core-}$ depending on whether the target galaxy is in the near or far side of the cluster, respectively.

The Category A galaxies used in Figs.~\ref{fig:1} and \ref{fig:3a} were obtained from the Key Project results\citep{free} that established the currently popular Hubble constant value of about 70 km~s$^{-1}$~Mpc$^{-1}$.  Examination of Fig.~\ref{fig:1} shows the appearance of two different types of distributions.  The data points for the galaxies in the sample with $z_\mathrm{H} <0.0024$ and $D_\mathrm{a} < 10$~Mpc (close sample galaxies) have a tight, linear appearance.  The data points for the galaxies in the sample with $z_\mathrm{H} >0.0024$ and $D_\mathrm{a} > 10$~Mpc (far sample galaxies) have a scattered, nearly unrelated appearance.  Figures~\ref{fig:1} and \ref{fig:3a} show that close sample galaxies, which are without an intervening cluster, have an effective $H_\mathrm{o} \approx 100$ km~s$^{-1}$~Mpc$^{-1}$ and a non-zero intercept with a correlation coefficient of 0.93 (see the Category C distance calculation in the Appendix).  The two outlier galaxies omitted in the Category C calibration have $D_\mathrm{a} > 11$~Mpc.  The far sample galaxies have at least one intervening cluster at 11~Mpc $ < x < 15$~Mpc and have a $D_\mathrm{a}$-$z_\mathrm{m}$ correlation coefficient of 0.53.  The SPM suggests this scatter is caused by $X_\mathrm{tc}$, $X_\mathrm{core}$, and $X_\mathrm{shell}$ of intervening galaxies.  A selection bias in selecting galaxies with Sink effects is expected because of the cluster cell structure.  Further, gravitational lensing observations have the $X$ factor effects of galaxies close to the light path.

The target galaxy effect $X_\mathrm{t}$ is caused by the Source or Sink in the target galaxy.  The $X_\mathrm{t}$ depends on the magnitude of the target galaxy like $X_\mathrm{mw}$.

\section{\label{sec:tiff}Discrete variations in redshift}

Beyond the local group, the major $X$ effects are $X_\mathrm{d}$, $X_\mathrm{shell}$, $X_\mathrm{tc}$, and $X_\mathrm{core}$.  The $X_\mathrm{source}$, $X_\mathrm{sink}$, and $X_\mathrm{g}$ effects are secondary and tend to partially offset each other.  At $x > 2$~Mpc the $X_\mathrm{d}$ is the major effect on $z$.  The major variation of $z_\mathrm{m}$ from the Hubble law is caused by the cluster factors $X_\mathrm{tc}$, $X_\mathrm{shell}$, and $X_\mathrm{core}$.  

\begin{figure}[!ht]
\begin{center}
\includegraphics[width=0.8\textwidth]{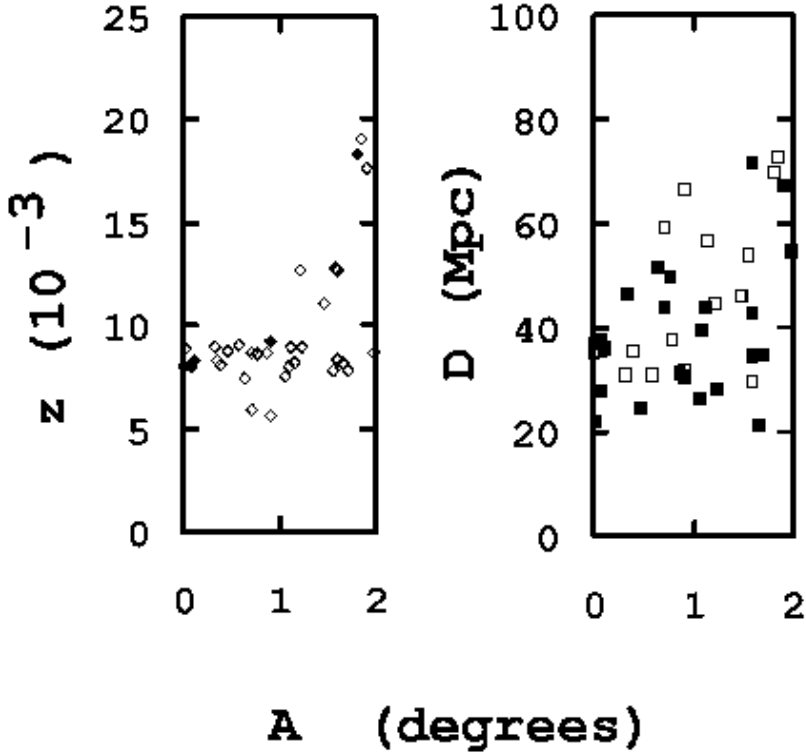}
\end{center}
\caption[width=3in]{ The left plot is of the measured redshift $z_\mathrm{m}$ versus the angle $A$ (arcdegrees) subtended from \astrobj{NGC~5353} (S0 in \astrobj{Canes Venatici}, $M = -20.8$ mag.) ($l,b,z$) = (82.61$^\circ$,71.63$^\circ$,8.0203$\times 10^{-3}$).  The open diamonds indicate the data points for Source galaxies.  The filled diamonds indicate the data points for Sink galaxies.  The right plot is the distance $D$ (Mpc) from earth versus $A$.  The open squares indicate the data points for galaxies with the $D$ calculated herein.  The filled squares indicate the data points for galaxies with the $D$ calculated using the Tully-Fisher relationship.}
\label{fig:1z}
\end{figure}

\begin{figure}[!hb]
\begin{center}
\includegraphics[width=0.8\textwidth]{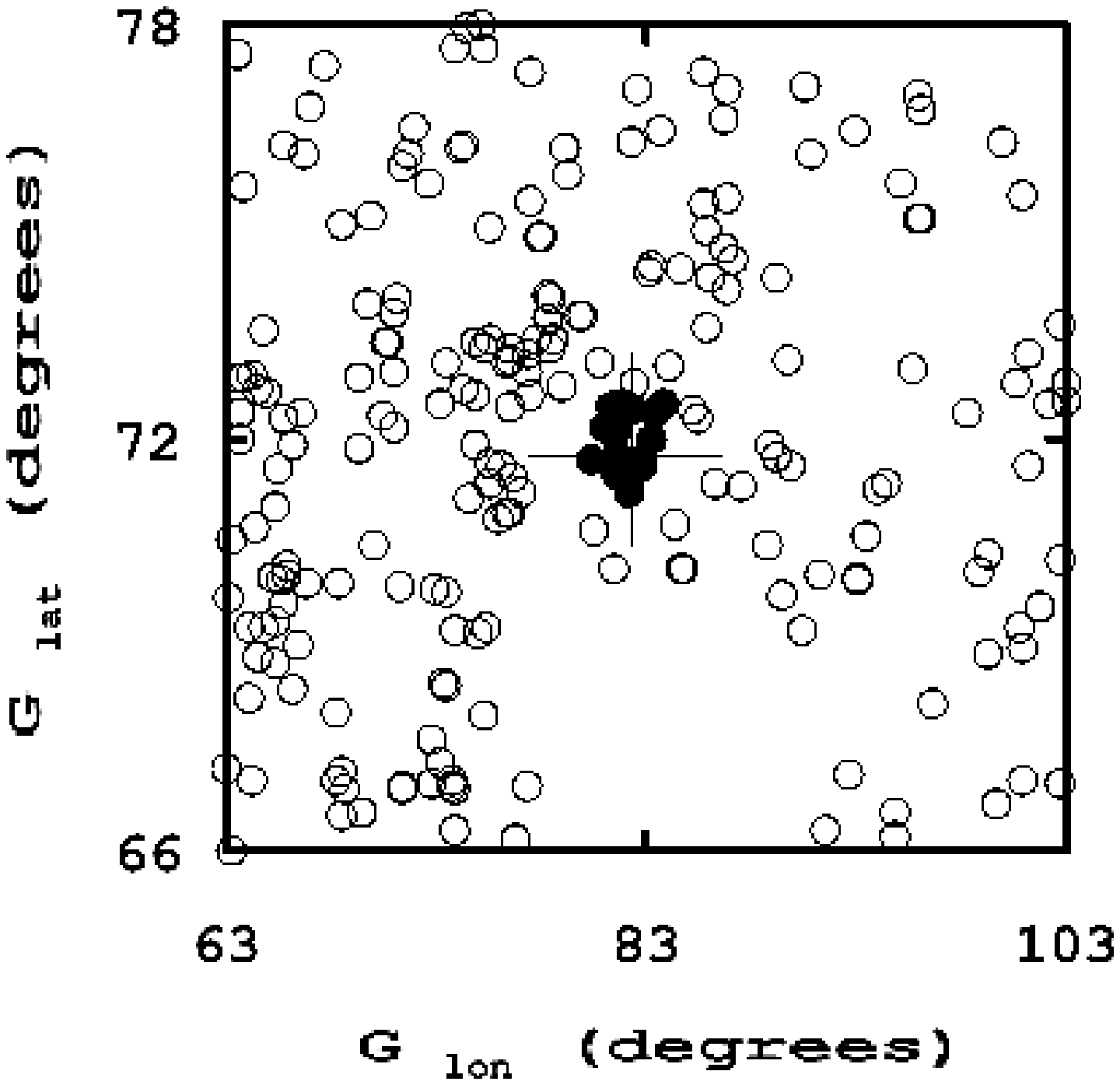}
\end{center}
\caption{Plot of the galactic latitude $G_\mathrm{lat}$ (arcdegrees) versus the galactic longitude $G_\mathrm{lon}$ (degrees) approximately six arcdegrees around \astrobj{NGC~5353}.  The open circles indicate the data points for galaxies more than one arcdegree from NGC~5353.  The filled circles indicate the data points for galaxies within one arcdegree of \astrobj{NGC~5353}.  The ``+'' or crosshairs indicate the position of \astrobj{NGC~5353}. }
\label{fig:2}
\end{figure}

\begin{figure}[!ht]
\begin{center}
\includegraphics[width=0.8\textwidth]{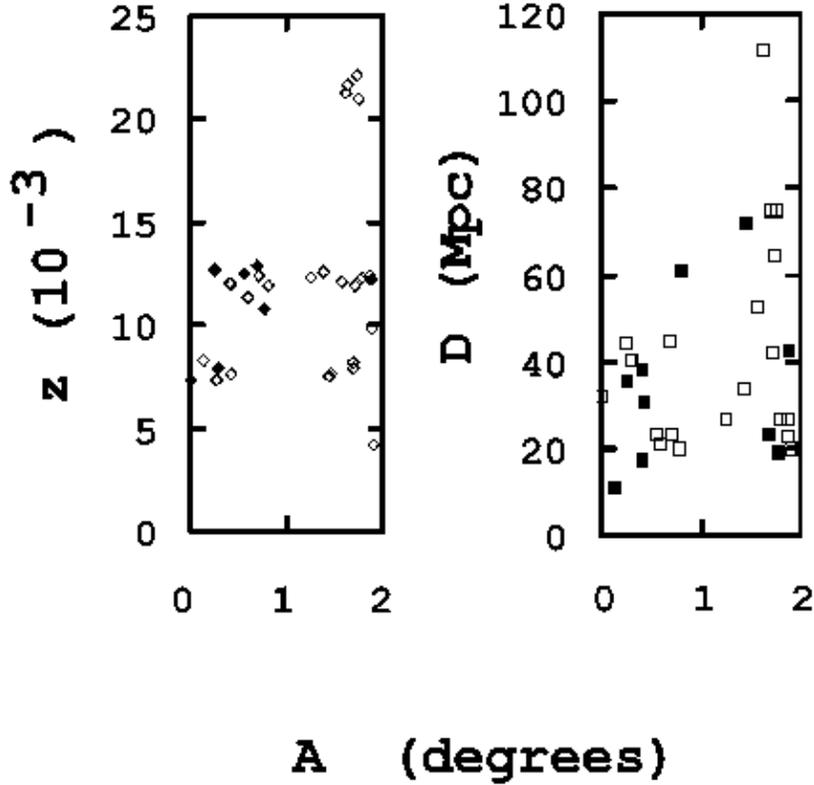}
\end{center}
\caption{The left plot is of the measured redshift $z_\mathrm{m}$ versus the angle $A$ (arcdegrees) subtended from \astrobj{NGC~2636} (E0 in \astrobj{Camelopardalis}, $M = -17.9$ mag.) ($l,b,z$) = (140.15$^\circ$,34.04$^\circ$,7.3163$\times 10^{-3}$).  The open diamonds indicate the data points for Source galaxies.  The filled diamonds indicate the data points for Sink galaxies.  The right plot is the distance $D$ (Mpc) from earth versus $A$.  The open squares indicate the data points for galaxies with the $D$ calculated herein.  The filled squares indicate the data points for galaxies with the $D$ calculated using the Tully-Fisher relationship.}
\label{fig:3}
\end{figure}

\begin{figure}[!hb]
\begin{center}
\includegraphics[width=0.8\textwidth]{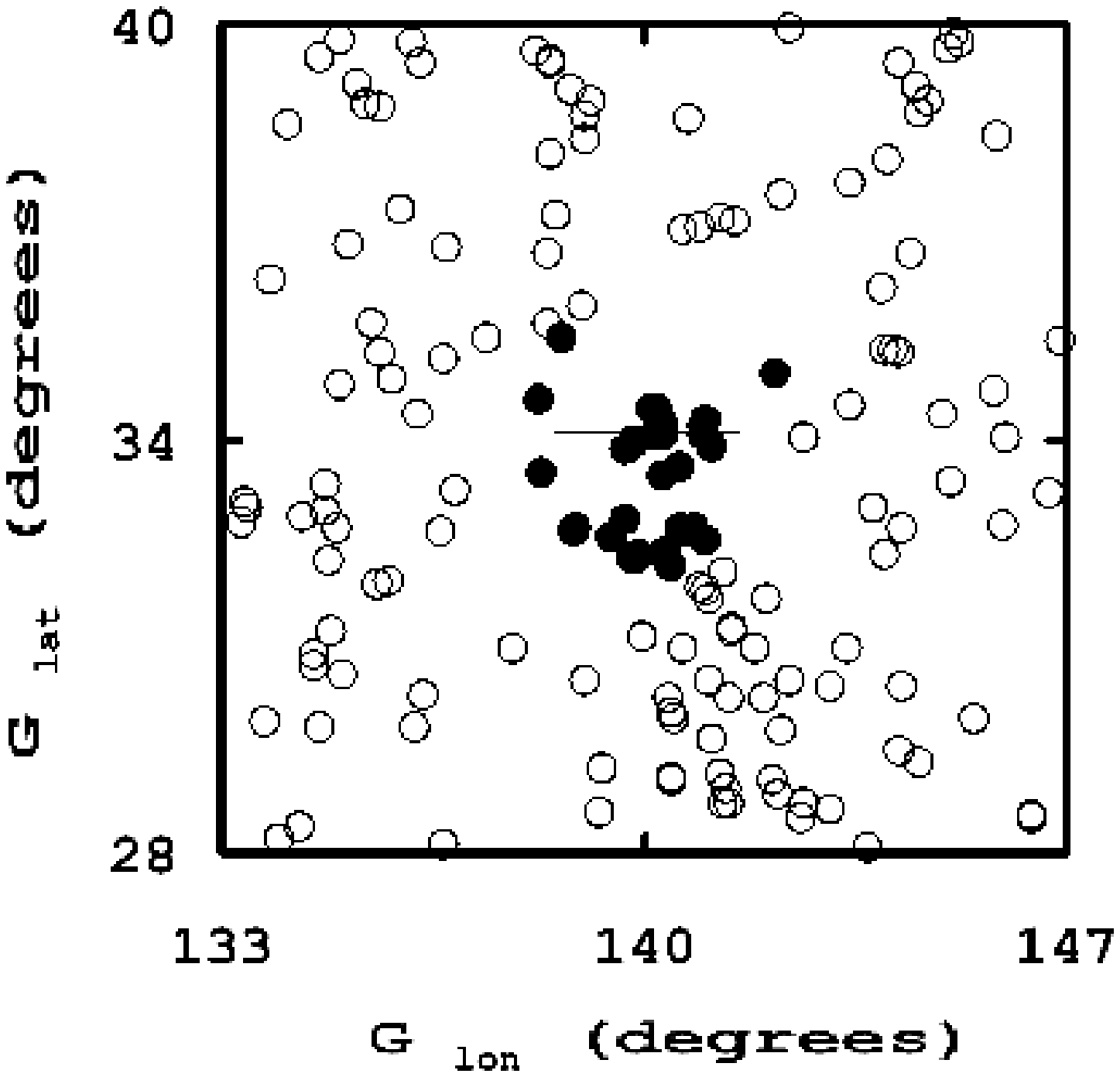}
\end{center}
\caption{Plot of the galactic latitude $G_\mathrm{lat}$ (arcdegrees) versus the galactic longitude $G_\mathrm{lon}$ (degrees) approximately six arcdegrees around \astrobj{NGC~2636}.  The open circles indicate the data points for galaxies more than one arcdegree from \astrobj{NGC~2636}.  The filled circles indicate the data points for galaxies within one arcdegree of \astrobj{NGC~2636}.  The ``+'' or crosshairs indicate the position of \astrobj{NGC~2636}. }
\label{fig:4}
\end{figure}

\begin{figure*}[!t]
\begin{center}
\includegraphics[width=\textwidth]{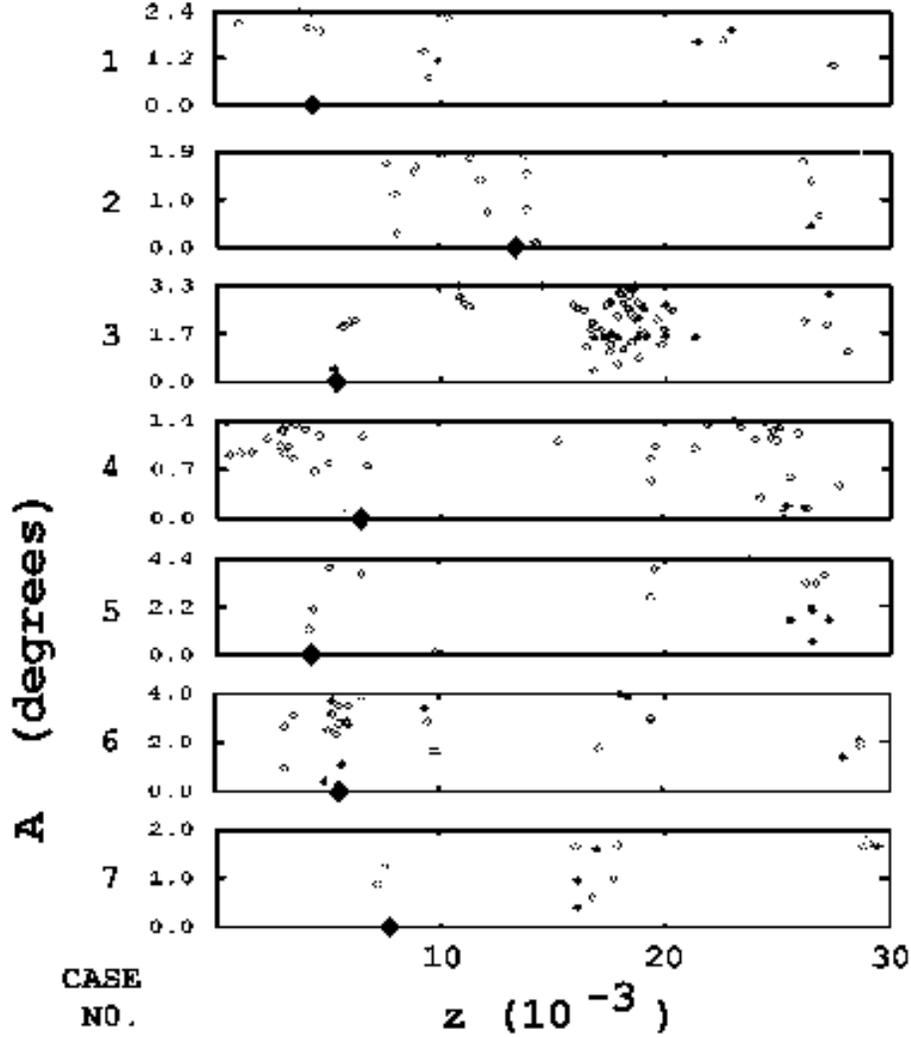}
\end{center}
\caption{Plots of the angle $A$ (arcdegrees) subtended from a target galaxy versus the measured redshift $z_\mathrm{m}$.  The target galaxy is shown on the $A=0$ axis as a large, filled diamond.  The open diamonds indicate the data points for Source galaxies.  The filled diamonds indicate the data points for Sink galaxies.  The data for the target galaxies are listed in Table~\ref{tab:5}.}
\label{fig:6z}
\end{figure*}

This effect is seen in Figs.~\ref{fig:1z}, \ref{fig:2}, \ref{fig:3}, \ref{fig:4}, and \ref{fig:6z}.  The left plots in Figs.~\ref{fig:1z} and \ref{fig:3} and the plots in Fig.~\ref{fig:6z} are $z$ versus $A$ from the target Sink galaxy.  The filled diamonds denote Sinks.  The right plots in Figs.~\ref{fig:1z} and \ref{fig:3} show $D$ of the galaxies versus $A$.  The filled squares are distances calculated using the Tully-Fisher relation \citep{tull77}.  The Category A galaxies (Cepheid galaxies) were used as calibration galaxies for the Tully-Fisher relation.  

%\begingroup
%\squeezetable
\begin{table*}[!t]
\scriptsize
\caption{\label{tab:5} The data for the Case Nos. of Fig.~\ref{fig:6z}. }
\begin{tabular}{clllrrlrll}
\hline
\hline
{Case}
&Constellation
&{Galaxy}
&{Morphology}
&$G_\mathrm{lon}$$^\mathrm{a}$
&{$G_\mathrm{lat}$$^\mathrm{a}$}
&{$D$} 
&{$z$}
&$M$
&Ratio \\
No.
&&&&$^\circ$
&$^\circ$
&Mpc
& $\times 10^{-3}$
&mag. \\
\hline
1&\astrobj{Camelopardalis}&\astrobj{UGC 05955}&E&135&42&13.8&4.2959&-16.68&1:2:5$^\mathrm{b}$\\
2&\astrobj{Canes}&\astrobj{NGC~5797}&S0/a&85&57&92.2&13.7083&-21.05&1:2$^\mathrm{c}$\\
3&\astrobj{Cetus}&\astrobj{NGC~0693}&S0/a?&148&-54&13.2&5.4335&-17.51&1:2:3:4:5$^\mathrm{d}$\\
4&\astrobj{Coma}&\astrobj{VCC 0546}&dE6&279&72&35.1&6.6833&-17.05&1:2$^\mathrm{c}$ \\
5&\astrobj{Fornax}&\astrobj{NGC~0802}&SAB(s)0+&294&48&17.2&4.4360&-17.50&1:2:4:5\\
6&\astrobj{Grus}&\astrobj{IC 5267B}&S0$\hat{ }$0$\hat{ }$&350&-62&12.0&5.5945&-17.74&1:2:4:5$^\mathrm{e}$\\
7&\astrobj{Hercules}&\astrobj{UGC 10086}&S0/a? &29&46&21.3&7.6864&-16.40&1:2:4$^\mathrm{f}$\\
\hline
\end{tabular}

$^\mathrm{a}${Rounded to the nearest degree. }
$^\mathrm{b}${The data point at $z \approx 27.5\times 10^{-3}$ is 6.4X.}
$^\mathrm{c}${Note the Source galaxies closer to earth. }
$^\mathrm{d}${Note the presence of a number of Sinks at $17\times 10^{-3} <z< 21\times 10^{-3}$ causing the lapse in the discrete ratio. }
$^\mathrm{e}${Note the presence of additional Sinks at 1x and 2x and the odd Source at approximately 3.5x.}
$^\mathrm{f}${Note the presence of additional Sinks at 2x causing the nearby Sources  to have lower $z$.}
\end{table*}
%\endgroup
%\end{minipage}
%\end{figure}

Figures~\ref{fig:2} and \ref{fig:4} show $G_\mathrm{lat}$ versus $G_\mathrm{lon}$ of the galaxies within approximately six arcdegrees surrounding the identified Sink.  The angular location of the identified Sink is marked by the crosshairs.  The filled circles denote the galaxies within one arcdegree of the identified Sink of Figs.~\ref{fig:1z} and \ref{fig:3}, respectively.  Source galaxies surround the Sink galaxies.

Figure~\ref{fig:1z} shows the $X_\mathrm{core}$ effect around the Sink \astrobj{NGC~5353}.  The $z$ value of galaxies closer than the identified Sink is increased due to $X_\mathrm{core+}$.  The $z$ value of galaxies farther than the identified Sink is decreased due to $X_\mathrm{core-}$.  The overall effect is the range of $z$ values of galaxies around the identified Sink are tightened toward the $z$ value of the identified Sink.

Figure~\ref{fig:3} shows the additional effect of $X_\mathrm{tc}$ in the $z$ values of more distant target galaxies that also have nearby Sinks and whose light must pass through another, closer cluster.  The nearer cluster causes an incremental increase in $z$ and the farther Sink in the cluster of the target galaxy causes the distribution of $z$ values to tighten.  Therefore, the total effect is of the two clusters.  If the $X_\mathrm{tc}$ effect of clusters are approximately equal, then the $z$ value of the farther group is $n X_\mathrm{tc}$, where $n$ is an integer count of the clusters the light passes through.  The net effect is a near scarcity of $z$ values half way between integer $z$ variation of the $z$ of the first cluster from earth.

\begin{figure}[!t]
\begin{center}
\includegraphics[width=0.8\textwidth]{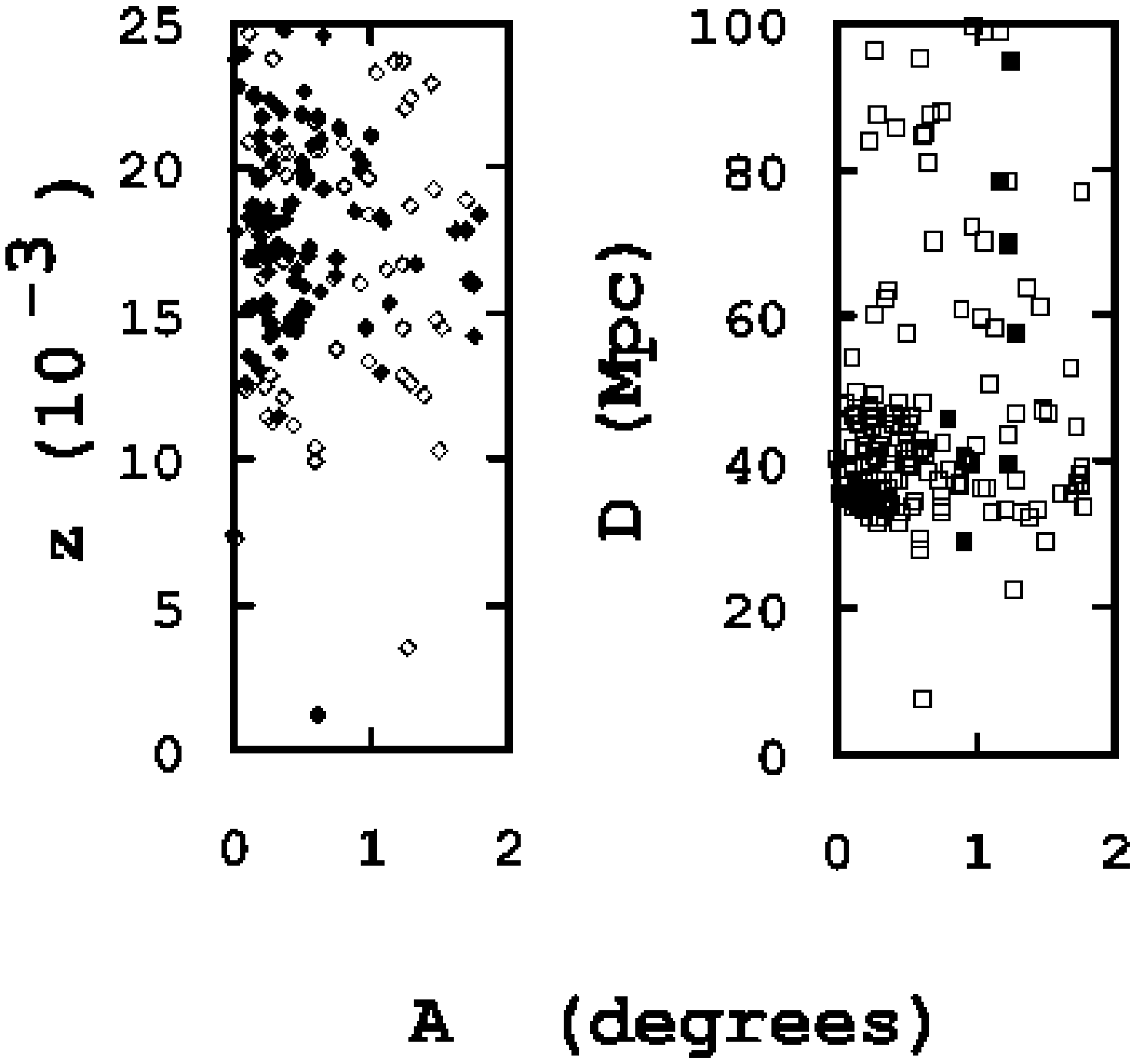}
\end{center}
\caption{The left plot is of the measured redshift $z_\mathrm{m}$ versus the angle $A$ (arcdegrees) subtended from \astrobj{NGC~1282} (E in \astrobj{Perseus}, $M = -19.8$ mag.) ($l,b,z$) = (150$^\circ$,-13.34$^\circ$,7.4727$\times 10^{-3}$).  The open diamonds indicate the data points for Source galaxies.  The filled diamonds indicate the data points for Sink galaxies.  The right plot is the distance $D$ (Mpc) from earth versus $A$.  The open squares indicate the data points for galaxies with the $D$ calculated herein.  The filled squares indicate the data points for galaxies with the $D$ calculated using the Tully-Fisher relationship.}
\label{fig:5}
\end{figure}

Figure~\ref{fig:5} of the \astrobj{Perseus} constellation shows the situation may become very complex.  The previous plots used situations with few Sink galaxies for clarity.

The calculation of $z$ from Eq.~(\ref{eq:26}) presents a formidable task.  Also, as distance increases, a bias develops resulting from the omission of undetected galaxies.  The effects of $X_\mathrm{d}$, $X_\mathrm{tc}$, $X_\mathrm{shell}$, and $X_\mathrm{core}$ relative to $X_\mathrm{source}$, $X_\mathrm{sink}$, and $X_\mathrm{g}$ suggest the $z$ may be calculated by adding the effects of the local group to the $z$ at $x<3$~Mpc, fitting a linear $D^{-1}$ relation to the $\rho - D$ curve for 3~Mpc$<x<8$~Mpc, applying the linear relation to the $\rho - D$ curve for $x>8$~Mpc, and modifying the $\rho$ calculation by counting the number of clusters the light passes near or through.  This method may be more appropriate than the Hubble Law for situations such as gravitational lensing in which the homogeneous assumption is invalid.

\section{\label{sec:disc}Discussion and conclusion}

The evidence is increasing that the discrete or periodic nature of redshift is genuine.  

\citet{milg} found the observations of $L^*$ elliptical galaxies \citep{roma} are in full concordance with the Modified Newtonian Dynamics model (MOND).  However, \citet{prad} using velocities of satellites and \citet{treu} using lensing and stellar dynamics of bright elliptical galaxies found results that contradict MOND.  If $F_\mathrm{s}$ and $F_\mathrm{g}$ of only the target galaxy act on a test particle in a galaxy, the $F_\mathrm{s}$ appears as a modification to $F_\mathrm{g}$.  The SPM allows the $F_\mathrm{s}$ of neighboring galaxies to also influence test particles.  The addition of $F_\mathrm{s}$ from neighboring galaxies, which is not radial in general, in a dense environment is unaccounted in MOND.  Therefore, because of the differing environmental densities, the SPM expects MOND to apply to $L^*$ elliptical galaxies and to fail for bright elliptical galaxies.  This suggests MOND may be an application of the SPM for low to medium galaxy densities and the modified acceleration of MOND is another force ($F_\mathrm{s}$).  

The calculation of $\rho$ involved only the luminosity and position of galaxies.  The SPM posits $\rho$ and the variation in $\rho$ is the major influence on photons redshift.  The model in this Paper considered gravitational forces or potentials as being linearly related to the $\epsilon$ or $\eta$ of the galaxies.  Therefore, as done herein, the SPM suggests the strength (luminosity) and position of galaxies are necessary and sufficient to determine intergalactic effects. 

The $F_\mathrm{s}$ derives from $\vec{\nabla} \rho$.  Therefore, the intergalactic $F_\mathrm{s}$ may be directed at any angle relative to our line of sight.  If the universe is finite and flat, then there is an edge where $\rho \rightarrow 0$ and where $\vec{\nabla} \rho \rightarrow 0$ is directed away from us.  Therefore, the objection to a Newtonian flat, finite universe is overcome. 

At larger distance (say $>200$~Mpc) the $F$ and $P$ terms approach an average per unit distance because the total number of encounters with Sources and Sinks per unit distance becomes nearly constant.  Therefore, if the volumes considered are greater than the spacing between several clusters, the mass distribution of the universe appears homogeneous and the Hubble Law is recovered for cosmological scale distance.  However, the error (peculiar velocity) in the Hubble Law's proportionality calculation is much larger than currently accepted.  Indeed, the constant of proportionality may vary with direction.  Examining the $P$ and $F$ data of the more distant sample galaxies herein suggests the error is at least $\partial z >0.016$, which is consistent with \citet{russ}. 

The purpose of the present investigation was to define the structure of a scalar potential field to be consistent with various galaxy observations.  The special focus was on observations that are inconsistent with other currently popular cosmological models.  This paper has not addressed the physical mechanism of Sources ($\epsilon$) and Sinks ($\eta$), the nature of $\rho$, or the nature of the $m_\mathrm{s}$ property of matter except to exclude the mass property.  However, this Paper has found a physical effect of these quantities.  A speculative example of a possible physical mechanism is galactic wind \citep{paum,shu,silk,veil}.  The shells of outward flowing, shocked gas around the Galactic center \citep[page 595]{binn}\citep{koni} and the presents of heavy-element ions in the intracluster medium suggests the emission of the ubiquitous, corporeal wind media is from the center of spiral galaxies.  If all stars emit matter in addition to radiation, Newton's spherical property and, hence, the inverse square law may be applied to the effect of the wind.  The gradient of the energy of the wind media ($\rho$) exerts a repulsive force ($F_\mathrm{s}$) on the cross-sectional area of particles ($m_\mathrm{s}$).  The rate of emission by Sources and of absorption by Sinks may be identified with $\epsilon$ and $\eta$, respectively.  The $G_\mathrm{s} m_\mathrm{s}/ G m_\mathrm{g}$ ratio of Eq.~(\ref{eq:49}) produce differing $F_\mathrm{s}$ for differing particle density.  Such a model is consistent with the radial distribution of black holes, quark stars, neutron stars, high metallicity stars, and lighter hydrogen stars.  In the outer bulge H$_\alpha$ RCs of varying metallicity, hot stars have peculiar shapes and differ from H{\scriptsize{I}} RCs of hydrogen gas.  However, in the outer disk H$_\alpha$ RCs of hot, hydrogen stars approach the H{\scriptsize{I}} RCs (see for example \citet{semp} for \astrobj{NGC~4321}).  

Because the Source of the $\epsilon$ is in the center of the Galaxy, the $\rho$ variation across our solar system is insignificant.  For $\delta r = 100$~AU, $ \delta(r^2)/r^2 \approx 10^{-17}$.  However, a gravitational experiment must carefully consider the differing $m_\mathrm{s}$ effects of particles.

The development of $X_\mathrm{d}$ in Section~\ref{sec:Xfactor} included a central Source and $\rho \propto r^{-1}$.  These conditions yielded the Hubble Law and $H_\mathrm{o}$.  In the part of a spiral galaxy where neighboring galaxies have an insignificant effect, the $\rho$ has the same, inverse square law form.  Therefore, the SPM suggests the appearance of $H_\mathrm{o}$ in galaxy observations is sufficient to indicate that an action of $F_\mathrm{s}$ is present and is insufficient to indicate that cosmological scale, general relativity effects are present.  The latter is consistent with the finding of \citet{coop} that the expansion of the universe is essentially negligible for scales up to galactic clusters.

For a sample of 32 galaxies with distances calculated using Cepheid variable stars, the redshift $z_\mathrm{c}$ calculated using the SPM $\rho$ variation had a 0.88 correlation coefficient to the measured redshift $z_\mathrm{m}$.  The calculation argued spiral galaxies are Sources and elliptical and other galaxies are Sinks.  Because the variation of $\rho$ appears to influence $z_\mathrm{m}$, the variation of $\rho$ was qualitatively related to the observed, discrete variation in $z_\mathrm{m}$. 

%\acknowledgments
\begin{ack}

This research has made use of the NASA/IPAC Extragalactic Database (NED) which is operated by the Jet Propulsion Laboratory, California Institute of Technology, under contract with the National Aeronautics and Space Administration.

This research has made use of the LEDA database (http://leda.univ-lyon1.fr).

I acknowledge the helpful comments by the reviewer.

I acknowledge and appreciate the financial support of Maynard Clark, Apollo Beach, Florida, while I was working on this project.

\end{ack}

\appendix

\section{\label{sec:initial}INITIAL DISTANCE ESTIMATION}
Category A galaxies consisted of 32 galaxies.  The distances $D_\mathrm{a}$ to Category A galaxies were calculated using Cepheid variable stars as listed by \citet{free} and \citet{macr}.

\begin{figure}
\includegraphics[width=0.8\textwidth]{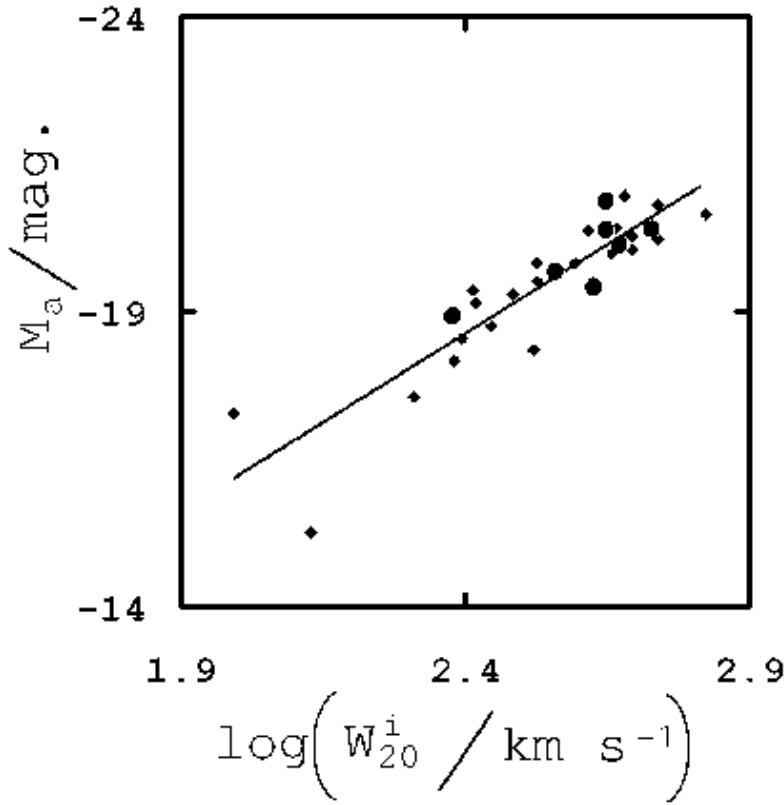}
\caption{Plot of the absolute magnitude $M_\mathrm{a}$ versus the inclination corrected 21 cm. line width $W^i_\mathrm{20}$ at 20\% of peak value for 29 of the 32 Category A galaxies \citep{free,macr}.  The straight line is a plot of Eq.~(\ref{eq:31}).  The circles indicate the data points for galaxies with (l,b) = (290$^\circ \pm 20^\circ$,75$^\circ \pm 15^\circ$).}
\label{fig:2a}
\end{figure}

Category B galaxies consisted of 5967 spiral galaxies in the sample that were not Category A galaxies with $W_\mathrm{20}$, $i_\mathrm{n}$, and $m_{b}$ values listed in the databases.  The distance $D_\mathrm{b}$ (Mpc) for each of the Category B galaxies were calculated following the method of Tully-Fisher \citep{tull77} as follows: (1) For the Category A galaxies, the absolute magnitude $M_\mathrm{a}$ was calculated,
\begin{equation}
\frac{M_\mathrm{a}}{\mathrm{mag.}}= \frac{m_{b}}{\mathrm{mag.}} - \frac{E_{\mathrm{xt}}}{\mathrm{mag.}} - 25 - 5 \, \log_{10} \left( \frac{D_\mathrm{a}}{\mathrm{Mpc}} \right)
\label{eq:30}.
\end{equation}
(2) A plot of $M_\mathrm{a}$/mag. versus $\log(W_\mathrm{20}^i/ \mathrm{km \, s}^{-1})$, where $W^i_\mathrm{20}$ is the inclination corrected $W_\mathrm{20}$, is presented in Fig.~\ref{fig:2a}.  IC 1613 (an Irregular galaxy classified as a Sink), NGC~5253 (an Irregular galaxy classified as a Sink), and NGC~5457 (\citet{free} noted the $D_\mathrm{a}$ was calculated differently) were omitted from the plot.  The straight line in Fig.~\ref{fig:2a} is a plot of 
\begin{equation}
\frac{M_\mathrm{a}}{\mathrm{mag.}}= K_\mathrm{ws} \, \log \left( \frac{W^i_\mathrm{20}}{\mathrm{km \, s}^{-1}} \right) + K_\mathrm{wi}
\label{eq:31},
\end{equation}
where $K_\mathrm{ws}= -6.0 \pm 0.6$, and $K_\mathrm{wi}=-4 \pm 1$ at one standard deviation ( 1$\sigma$).  The correlation coefficient is -0.90.  \citet{tull77} calculated a similar relation with $K_\mathrm{ws}= -6.25$ and $K_\mathrm{wi}=-3.5$.  The circles indicate the data points for galaxies with (l,b) = (290$^\circ \pm 20^\circ$,75$^\circ \pm 15^\circ$) as in Fig.~\ref{fig:1}.  However, unlike Fig.~\ref{fig:1}, the data in Fig.~\ref{fig:2a} for the outlier galaxies appears consistent with the other sample galaxies' data.

(3) The $D_\mathrm{b}$ is 
\begin{equation}
D_\mathrm{b}= 10^{0.4 (m_{b} - M_\mathrm{b} - E_{\mathrm{xt}})}
\label{eq:33},
\end{equation}
where 
\begin{equation}
\frac{M_\mathrm{b}}{\mathrm{mag.}}= K_\mathrm{ws} \, \log \left(\frac{W^i_\mathrm{20}}{\mathrm{km \, s}^{-1}} \right) + K_\mathrm{wi}
\label{eq:33a}.
\end{equation}

\begin{figure}
\includegraphics[width=0.8\textwidth]{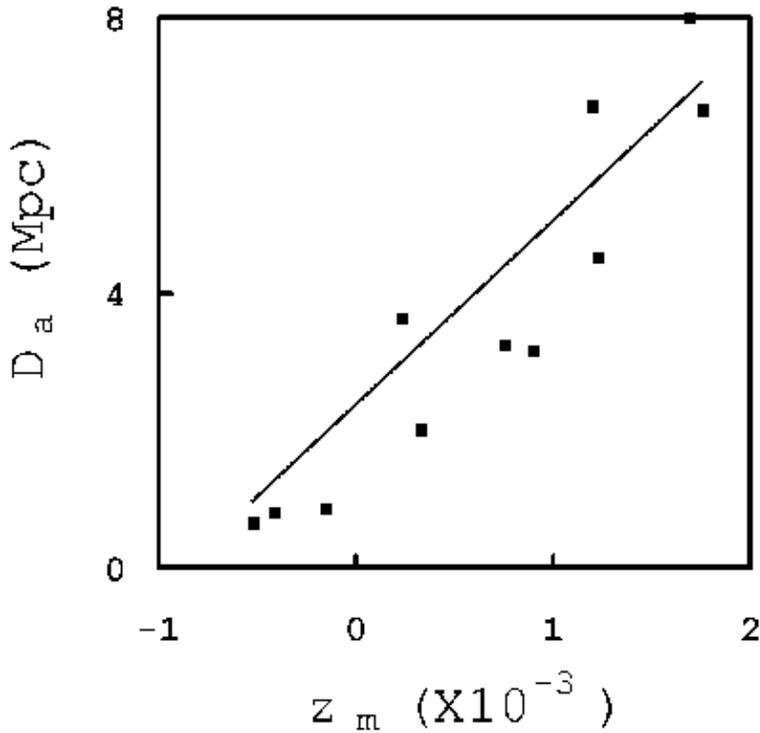}
\caption{Plot of distances $D_\mathrm{a}$ to galaxies calculated using Cepheid variable stars as listed by \citet{free} and \citet{macr} versus the redshift $ z_\mathrm{m}$ of these galaxies.  The straight line is a plot of Eq.~(\ref{eq:34}).}
\label{fig:3a}
\end{figure}

Category C galaxies consisted of galaxies in the sample that were not in the previous categories with  $-0.001  < z_\mathrm{m} <.002$.  A plot of $D_\mathrm{a}$ versus $ z_\mathrm{m}$ for the Category A galaxies in the Category C range is presented in Fig.~\ref{fig:3a}.  Because the goal of this section is to arrive at the initial distance estimation, NGC~2541 and NGC~4548 are outlier galaxies and were omitted from the plot.  The straight line in Fig.~\ref{fig:3a} is a plot of 
\begin{equation}
D_\mathrm{a} = \frac{ c z_\mathrm{m}}{ K_\mathrm{czs} } + K_\mathrm{czi}
\label{eq:34},
\end{equation}
where $K_\mathrm{czs}= 100 \pm 10 $ km~s$^{-1}$~Mpc$^{-1}$, and $K_\mathrm{czi}= 1.7 \pm 0.4$~Mpc at 1$\sigma$.  The correlation coefficient is -0.93. 

The distance $D_\mathrm{c}$ (Mpc) for Category C galaxies is 
\begin{equation}
D_\mathrm{c} = \frac{ c z_\mathrm{m}}{ K_\mathrm{czs}} + K_\mathrm{czi}
\label{eq:35a}.
\end{equation}

Category D galaxies are galaxies in the sample not in the previous categories and with $ z_\mathrm{m} < -0.001 $.  The distance $D_\mathrm{d}$ (Mpc) to Category D galaxies was 1~Mpc.

Category E galaxies are all other galaxies in the sample not in the previous categories.  The distance $D_\mathrm{e}$ (Mpc) to these galaxies was calculated using the Hubble law with $H_\mathrm{o}= 70$ km~s$^{-1}$~Mpc$^{-1}$.

% The phrase \citep{Bai92} produces (Bailyn 1992).
% In the phrase \citeasnoun{Bai95} Bailyn et al. (1995) appear as a noun.
% Affixes (e.g. Barnes et al. 1976) are produced by the phrase
% \citeaffixed{Barnes et al. 1976}{e.g.}.
% Other options of the harvard package, e.g. \citeyear, are not
% reproduced in New Astronomy.

%\begin{thebibliography}{}

% \harvarditem{Name}{Year}{label}
% Text of bibliographic item
%\harvarditem{Author}{year}{label}Reference

%\end{thebibliography}

\end{document}